\documentclass[11pt]{article}

\usepackage{amsmath}
\usepackage{amssymb}
\usepackage{setspace}
\usepackage{bbm}
\usepackage{dsfont}
\usepackage{graphicx}
\usepackage{pict2e}

\def\P{\mathbb{P}}
\def\C{\mathbb{C}}
\def\Z{\mathbb{Z}}

\def\tr{\operatorname{tr}}
\def\Hom{\operatorname{Hom}}
\def\codim{\operatorname{codim}}
\def\dag{\dagger}
\def\rank{\operatorname{rank}}
\def\ker{\operatorname{ker}}

\def\cA{\mathcal{A}}

\def\cE{\mathcal{E}}
\def\cF{\mathcal{F}}

\def\cL{\mathcal{L}}
\def\cU{\mathcal{U}}

\def\cO{\mathcal{O}}
\def\cX{\mathcal{X}}

\def\cQ{\mathcal{Q}}

\def\2pii{(2 \pi i)}

\def\sig{\sigma}
\def\Sig{\Sigma}
\def\sigb{\bar{\sigma}}

\def\tW{\widetilde{W}}
\def\tm{\widetilde{m}}

\def\Phib{\bar{\Phi}}

\def\ie{i.e., }
\def\kd{{k_d}}
\def\cC{\mathcal{C}}
\def\hx{\hat{x}}
\def\hp{\hat{p}}
\def\tx{\tilde{x}}
\def\ty{\tilde{y}}
\def\tX{\widetilde{X}}
\def\tY{\widetilde{Y}}
\def\cV{\mathcal{V}}
\def\tp{\tilde{p}}
\def\tP{\widetilde{P}}
\def\Sigh{\widehat{\Sigma}}
\def\sigh{\hat{\sigma}}
\def\phih{\hat{\phi}}

\def\phic{{\check{\phi}}}
\def\ac{{\check{a}}}

\newcommand{\be}{\begin{equation}}
\newcommand{\ee}{\end{equation}}
\newcommand{\bes}{\begin{equation*}}
\newcommand{\ees}{\end{equation*}}
\newcommand{\bea}{\begin{eqnarray}}
\newcommand{\eea}{\end{eqnarray}}
\newcommand{\beas}{\begin{eqnarray*}}
\newcommand{\eeas}{\end{eqnarray*}}

\newcommand{\bb}{\mathbb}
\newcommand{\p}{\partial}
\newcommand{\bin}[2]{\begin{pmatrix} #1 \\ #2 \end{pmatrix}}
\newcommand{\vev}[1]{\langle #1 \rangle}
\newcommand{\fl}[1]{{\lfloor #1 \rfloor}}

\begin{document}
\numberwithin{equation}{section}
{
\begin{titlepage}
\begin{center}
\hfill BONN-TH-2012-09\\
\hfill NSF-KITP-12-070\\
\hfill UCSB Math  2012-17\\
\vskip 0.75in

{\Large \bf Nonabelian 2D Gauge Theories for Determinantal\\ Calabi--Yau Varieties}\\

\vskip 0.4in

{ Hans Jockers${}^{a}$, Vijay Kumar${}^{b}$, Joshua M.~Lapan${}^{c}$, David R.~Morrison${}^{d,e}$, Mauricio Romo${}^{e}$}\\

\vskip 0.3in
{\small
\begin{tabular}{ll}
${}^{\, a}${\em Bethe Center for Theoretical Physics,} & ${}^{\, b}${\em Kavli Institute for Theoretical Physics,}\\
$\phantom{{}^{\, a}}${\em Physikalisches Institut, Universit\"at Bonn,} & $\phantom{{}^{\, b}}${\em University of California}\\
$\phantom{{}^{\, a}}${\em 53115 Bonn, Germany} & $\phantom{{}^{\, b}}${\em Santa Barbara, CA 93106, USA}\\[2ex]
\end{tabular}
\begin{tabular}{lll}
${}^{\, c}${\em Department of Physics,} 	& ${}^{\, d}${\em Department of Mathematics,}  	& ${}^{\, e}${\em Department of Physics,} \\
$\phantom{{}^{\, c}}${\em McGill University,} 				&$\phantom{{}^{\, d}}${\em University of California} 				& $\phantom{{}^{\, e}}${\em University of California} \\
$\phantom{{}^{\, c}}${\em Montr\'eal, QC, Canada}			&$\phantom{{}^{\, d}}${\em Santa Barbara, CA 93106, USA} 		&  $\phantom{{}^{\, e}}${\em  Santa Barbara, CA 93106, USA}\\[2ex]
\end{tabular}
}
\end{center}

\vskip 0.35in

\begin{center} {\bf Abstract} \end{center}

The two-dimensional supersymmetric gauged linear sigma model (GLSM) with abelian gauge groups and matter fields has provided many insights into string theory on Calabi--Yau manifolds of a certain type: complete intersections in toric varieties. 
In this paper, we consider two GLSM constructions with nonabelian gauge groups and charged matter whose infrared CFTs correspond to string propagation on determinantal Calabi--Yau varieties, furnishing another broad class of Calabi--Yau geometries in addition to complete intersections.  We show that these two models --- which we refer to as the PAX and the PAXY model --- are dual descriptions of the same low-energy physics. Using GLSM techniques, we determine the quantum K\"ahler moduli space of these varieties and find no disagreement with existing results in the literature.

\vfill

\noindent May 14, 2012

\let\thefootnote\relax\footnotetext{jockers@uni-bonn.de, vijayk@kitp.ucsb.edu, jlapan@physics.mcgill.ca, drm@math.ucsb.edu, mromo@physics.ucsb.edu}

\end{titlepage}
}

\newpage
\section{Introduction and Summary of Results}
\label{sec:overview}

The gauged linear sigma model (GLSM) is a tool to understand the $\mathcal{N}=2$ superconformal field theories that arise as the low-energy fixed points of supersymmetric nonlinear sigma models with Calabi--Yau target spaces \cite{Witten:1993yc}. The GLSMs that we study are two-dimensional $\mathcal{N}=(2,2)$ supersymmetric gauge theories with matter. The advantage of the GLSM approach is that it involves a free UV fixed point where the fields have linear couplings, as opposed to the nonlinear couplings of the nonlinear sigma model.  Some of the geometric quantities that describe the Calabi--Yau manifold, e.g., certain complex structure or K\"ahler moduli, can be interpreted in terms of couplings that appear in the GLSM Lagrangian. 

The familiar abelian GLSM was tailor-made for describing smooth complete intersection Calabi--Yau varieties embedded in an ambient toric variety. Let $V$ denote an ambient variety of dimension $D$, which is described by $D+s$ homogeneous coordinates $\phi_a$ that transform under a multiplicative $(\mathbb{C}^\star)^s$ action
\be
\phi_a \longrightarrow  \big({\textstyle\prod_{\ell=1}^s \lambda_\ell^{Q_a^\ell}}\big) \phi_a \, ,
\ee
and by a subset $F$ of the $\bb{C}^{D+s}$ spanned by the $\phi_a$.  In common notation \cite{Witten:1993yc,summing}, we denote these by
\be
V = \frac{\big( \bb{C}^{D+s} \ \backslash \ F\big)}{(\bb{C}^\star)^s} \, , \qquad  X := \big\{ \phi\in V \ \big| \ J^i(\phi) = 0 \, , ~ i=1,\ldots,r \ \big\} \, .
\ee
A smooth complete intersection Calabi--Yau $X \subset V$ is defined as the vanishing locus of $r$ quasi-homogeneous polynomials $J^i(\phi)$ that are transverse, \ie the $r\times (D+s)$ Jacobian matrix
\begin{equation}
\left. \frac{\partial J^i(\phi)}{\partial \phi_a}\, \right\vert_{\phi\in X}\, , \label{eq:jac-cond}
\end{equation}
has rank $r$ at every point of $X$.

In the GLSM, the homogeneous coordinates are the lowest components of $(2,2)$ chiral superfields, $\Phi_a$, while the $(\bb{C}^\star)^s$ action 
is implemented through a  $U(1)^s$ gauge symmetry under which the $\Phi_a$ carry charges $Q_a^\ell$, $\ell = 1, \cdots, s$ (the gauge symmetry action is effectively complexified in $\mathcal{N}=2$ theories). The Fayet-Iliopoulos (FI) parameters and theta angles of the gauge theory can be collected into $s$ complex parameters whose values correspond to the complexified K\"ahler moduli of $V$ when they lie inside the K\"ahler cone. Thus, this theory naturally encodes all the details of the ambient toric variety, $V$.  To describe a complete intersection $X \subset V$ of the quasi-homogeneous polynomials $\{J^{i=1,\cdots,r}\}$, one introduces $r$ additional chiral superfields $P_i$ and deforms the theory with the superpotential
\be
\label{eqn:abel-spot}
W = \sum_i P_i J^i(\Phi) \, .
\ee
The $P_i$ serve as Lagrange multipliers and their gauge charges are fixed uniquely by demanding the gauge invariance of the superpotential.
When the FI parameters are in the K\"ahler cone of $V$, the F-terms for the $P_i$ impose the vanishing conditions $J^i(\phi) = 0$ while the F-terms for the $\Phi_a$ and the Jacobian condition \eqref{eq:jac-cond} force the $p_i$ to vanish, thus yielding $X$ for the moduli space of supersymmetric vacua.

While the abelian GLSM is well-suited for describing complete intersections in toric varieties, these likely form only a small fraction of Calabi--Yau threefolds.  For a complete intersection $X \subset V$, the minimal generating set for the ideal $I(X)$ is composed of $r$ elements, where $r$ is the (complex) codimension of $X\subset V$. When $X$ is not a complete intersection --- which is the general situation when the codimension of $X$ is greater than one --- any set of generators of $I(X)$ will have more than $r$ elements, but these generators will satisfy non-trivial relations (syzygies). 
The mathematics literature contains several interesting constructions of 
non-complete-intersection Calabi--Yau threefolds motivated by results from the
algebra of syzygies.
For example, it is known that any threefold in $\P^5$ whose ideal is suitably ``regular'' and whose canonical bundle is the restriction of a line bundle from $\P^5$ is a complete intersection \cite{serre-codim-2}. However, a threefold in $\P^6$ with a suitably regular ideal (and canonical bundle arising by restriction) is instead
a Pfaffian determinantal
variety, \ie it is specified by the vanishing of the $2k\times 2k$ Pfaffians of a $(2k+1) \times (2k+1)$ skew-symmetric matrix \cite{buchsbaum1977algebra, okonek1994notes,walter1996pfaffian, Tonoli2006}. When $k > 1$, $X$ is not a complete intersection: the rank $2k-2$ locus is specified by the vanishing of all the $2k+1$ diagonal $2k\times 2k$ Pfaffian minors, while the codimension is $3 < 2k+1$. Tonoli has constructed a number of examples of Calabi--Yau threefolds in $\P^6$, and his work confirms the intuition that many more Calabi--Yau manifolds are non-complete intersections than complete intersections \cite{Tonoli2006}.
For a threefold in $\P^7$ with a suitably regular ideal, general results are not known but interesting examples are provided by a construction of Kustin--Miller \cite{kustinmiller} and by a construction of Gulliksen and Neg{\aa}rd \cite{gulliksen1971}; the latter construction is a special case of the determinantal varieties considered in this paper.

For a general non-complete intersection Calabi--Yau, we do not know how to build a GLSM taking into account the syzygies. To see the difficulty, suppose that $X$ is not a complete intersection and consider adding fields $P_i$ as we do for complete intersections.  By assumption, there exists at least one set of non-trivial polynomials $F_i(\phi) \in \frac{\mathbb{C}[\phi]}{I(X)}$ for which $\sum_i F_i(\phi) J^i(\phi) = 0$.  This means that the superpotential (\ref{eqn:abel-spot}) is invariant under the transformation $P_i\rightarrow P_i+P_0 F_i(\Phi)$, so the degree of freedom corresponding to $P_0$ will be unconstrained and result in a moduli space that contains a line bundle (with section $p_0$) over $X$, rather than just $X$ which was our aim.  In general, having chosen a basis $J^i(\phi)$ for $I(X)$ with $d > r$ elements, we would wind up with a rank $(d-r)$ bundle over $X$.

While building a GLSM for a general Calabi--Yau subvariety seems difficult, there is a simple class of non-complete intersections called ``determinantal varieties'' for which it is possible.  A determinantal variety is defined as follows: first, consider homogeneous coordinates $\phi_a$ on an ambient toric variety $V$ and define an $m$ by $n$ matrix $A(\phi)$ with entries given by quasi-homogeneous polynomials in $\phi$.  A determinantal variety is then defined by the locus in $V$ where $A(\phi)$ degenerates to rank $k$ or below, which we denote by $Z(A,k)$. 
The  corresponding ideal $I(Z(A,k))$ is generated by the $(k+1)\times (k+1)$ minors of $A(\phi)$ \cite{Harris1995}.  A determinantal variety is generally not a complete intersection since $I(Z(A,k))$ is generated by ${m \choose k+1}{n \choose k+1}$ polynomials, while the codimension of $Z(A,k) \subset V$ is $(m-k)(n-k)$.\footnote{A general $m\times n$ matrix $A(\phi)$ has $mn$ degrees of freedom. On the locus where $A(\phi)$ degenerates to rank $k$, there are $k$ linearly independent rows yielding $kn$ degrees of freedom, while the remaining $(m-k)$ rows are linear combinations of the first $k$, contributing $k(m-k)$ degrees of freedom.  Thus, along $Z(A,k)$ the degrees of freedom in $A(\phi)$ are reduced by $mn-nk-k(m-k) = (m-k)(n-k)$, which is the codimension.}

We were led to this line of investigation by the work of Hori and Tong \cite{Hori:2006dk} who constructed a GLSM to provide a ``physics proof'' of R{\o}dland's conjecture that a certain Pfaffian Calabi--Yau in $\P^6$ is in the same quantum K\"ahler moduli space as a complete intersection in the Grassmannian $G(2,7)$ \cite{Rodland9801}. The GLSM they constructed showed that the moduli space has two large volume points, corresponding to the two Calabi--Yau manifolds, and three other singular points, exactly as predicted by R{\o}dland. The analysis was complicated due to the presence of a nonabelian $U(2)$ gauge symmetry. 
GLSMs with nonabelian gauge symmetry have been studied in the literature, particularly in the context of understanding complete intersections in Grassmannians \cite{Witten:1993yc, Witten:1993xi, Hori:2006dk, Lerche:2001vj} and in partial flag manifolds \cite{Donagi:2007hi}, and more recently in the context of understanding (skew-symmetric and symmetric) determinantal varieties \cite{Hori:2006dk, Donagi:2007hi, Hori:2011pd}.
We are also aware of some interesting
work in progress \cite{HoriKnappLectures} studying
the phase structure of skew-symmetric determinantal varieties.

In this paper we construct GLSMs that describe determinantal varieties whose defining matrices have no symmetry properties. These models have a gauge group with a unitary group factor and charged matter in the fundamental and anti-fundamental representation. The outline of the paper is as follows.  For the remainder of this section, we present an overview and summary of results.  In Section 2, we discuss some basic properties of determinantal varieties and their desingularizations (known as incidence
correspondences) that are relevant to the GLSM. In Section 3, we discuss two ways in which a rank constraint may be imposed on a matrix,
leading to two GLSMs: the ``PAX'' model and the ``PAXY'' model, which generalize the orthogonal and symplectic versions studied in \cite{Hori:2006dk, Hori:2011pd} to unitary groups. We also show that the two models are dual descriptions of the same low-energy physics.  In Section 4, we analyze in detail the phase structure of linear determinantal varieties, and in Section 5 we apply our analysis to examples of determinantal Calabi--Yau threefolds that have appeared in the literature \cite{Bertin0701,Kapustka0802, Hosono:2011vd}. We end with conclusions and directions for future work in Section~6.

\subsection{Nonabelian GLSMs for Determinantal Varieties: General Idea}

The question in the context of the GLSM is how to impose a rank condition on $A(\phi)$.  We will explore two classes of nonabelian GLSMs, which we will call the PAX and PAXY models, that turn out to be equivalent under a two-dimensional version of Seiberg duality \cite{Seiberg:1994pq} explained by Hori \cite{Hori:2011pd}.  These two classes of nonabelian GLSMs generalize those studied in great detail in \cite{Hori:2006dk,Hori:2011pd} to unitary groups.  For reasons discussed in Section \ref{sec:generalize}, we focus on square matrices $A(\phi)$, so $m=n$ from now on unless stated otherwise.  The ambient variety $V$ is defined in the usual way, through a $\mathcal{N}=(2,2)$ $U(1)^s$ gauge theory with complexified FI parameters $t_\ell$  and chiral superfields $\Phi_a$ with charges $Q_a^\ell$, where $a=1,\ldots,D+s$, and $\ell=1,\ldots, s$.  Obtaining a determinantal subvariety of $V$ requires adding matter and turning on a superpotential:

\paragraph{PAX model:}~\\
To impose the condition $\rank(A(\phi)) \leq k$, one strategy is to demand that a rank $n-k$ matrix lie in the kernel of $A(\phi)$. Let $x$ be a $n\times \kd$ matrix of rank $\kd$ with 
\begin{equation}
  \kd := n-k \ .
\end{equation}  
The condition we wish to enforce,
\begin{equation}
A(\phi)\, x = 0\, , \label{eq:rank-kernel}
\end{equation}
is preserved by a $GL(\kd,\C)$ action on the columns of $x$. We can impose this condition through the superpotential
\begin{eqnarray}
\label{eqn:PAX}
W=\sum_{i,j=1}^{n}\sum_{\alpha=1}^{n-k}P_{\alpha i}A(\Phi)_{ij}X_{j\alpha}=\mathrm{tr}\left(PAX\right)\, ,
\end{eqnarray}
and a gauge symmetry $U(\kd)$, where $x_{j}$ transform as anti-fundamentals and the $p_{i}$, which act as as Lagrange multipliers, transform as fundamentals.  The $U(1)^s$ charges must be assigned so that the superpotential is invariant and so that there is no axial anomaly --- similarly for $\det(U(\kd))$ charges. 
Each row of $X$ (and column of $P$) must have definite $U(1)^s$ charges, otherwise the $U(\kd)$ symmetry would be broken.  (We could equivalently regard the $x_j$ as the Lagrange multipliers imposing $p\, A(\phi) = 0$; the distinction between the two will depend on the phase determined by the FI parameters.)  The various F-terms and D-terms place further constraints on the model that depend on the FI parameters; this phase structure will be discussed in detail in Section \ref{sec:PAX}.

\paragraph{PAXY model:}~\\
Alternatively, note that along $Z(A,k)$ the matrix $A(\phi)$ can be factorized into a product of two matrices $\tx$ and $\ty$ of dimensions $k\times n$ and $n\times k$, respectively,
\begin{equation}
A(\phi) = \ty \, \tx\, . \label{eq:rank-factor}
\end{equation}
If $\rank(\ty) = k$, then $\rank(A(\phi)) = \rank(\tx)$ and this decomposition is unique up to $GL(k,\C)$ transformations
\begin{equation}
\tx \rightarrow M \tx, \qquad \ty \rightarrow \ty M^{-1}\, , \qquad M \in GL(k,\C)\,
\end{equation}
(and similarly if $\rank(\tx)=k$ while $\rank(\ty) \leq k$).  Since $\mathfrak{gl}(k,\mathbb{C}) = \mathfrak{u}(k)_\mathbb{C}$, this suggests incorporating a $U(k)$ gauge symmetry into the GLSM with $n$ fundamentals $\tX_i$ and $n$ anti-fundamentals $\tY_i$. The factorization condition \eqref{eq:rank-factor} can then be imposed by introducing a $n\times n$ matrix of chiral superfields $\tP$ that serve as Lagrange multipliers.  The superpotential is
\be
\label{eqn:nonabel-spot}
W  =  \sum_{i,j=1}^{n} \tP_{ji} \Big( A(\Phi)_{ij} - \sum_{\hat{\alpha}=1}^k \widetilde{Y}_{ i\hat{\alpha}} \widetilde{X}_{\hat{\alpha} j} \Big) = \textrm{tr}\Big\{ \tP \big( A(\Phi) - \widetilde{Y} \widetilde{X} \big) \Big\}  \, .
\ee
Again, the $U(1)^s$ and $\det(U(k))$ charges must be chosen so that the superpotential is invariant and so that there is no axial anomaly.  Furthermore, each row of $\widetilde{Y}$ (and column of $\widetilde{X}$) must have a definite charge under $U(1)^s$, otherwise we would break the $U(k)$ symmetry.

A variant on the above models would be to require $A(\phi)$ to be symmetric or antisymmetric, yielding orthogonal or symplectic groups, respectively, as discussed briefly in Section \ref{sec:generalize}.  The PAX version of the symplectic model was used by Hori and Tong to study the R{\o}dland Calabi--Yau: the rank four locus of a $7\times 7$ antisymmetric matrix of linear entries in $\P^6$ \cite{Hori:2006dk}. The PAXY model was introduced by Hori in \cite{Hori:2011pd} as another description of the R{\o}dland example, and the orthogonal models were also extensively studied.  As explained there, the PAXY and PAX models are dual to each other in a way analogous to Seiberg duality in four dimensions \cite{Seiberg:1994pq}, \ie they flow to the same $(2,2)$ IR superconformal field theory.

\subsection{Summary of Results}
The vacuum moduli space of the PAX and PAXY models has the geometry of a resolved determinantal variety, which we will call $X_A$ and $\hat{X}_A$, known as an incidence correspondence.  We demonstrate that the PAX and PAXY models are dual descriptions of the same low-energy physics by performing various checks.  For one, we show that in a certain regime of parameter space (near large volume points), both models flow to IR superconformal field theories associated with distinct but isomorphic vacuum moduli spaces --- since the FI parameters correspond to marginal deformations of the superconformal field theory, this duality should extend to the whole moduli space.  We also verify that anomalies and central charges match in both models. We then focus our study largely on the PAX model since it has fewer fields and is in that sense simpler, though we make connections at various points with details of the PAXY model. 

The PAX model has a total of $s+1$ complexified FI parameters, and we can always find a phase (region of FI parameter space) corresponding to a large volume point describing the incidence correspondence $X_A$ as a resolution of the determinantal variety $Z(A,k)\subset V$.  
In general, the complete phase structure of the model is difficult to analyze, so we specialize to the simple class of linear determinantal varieties in $\P^D$ and determine the phase structure as a function of the complexified FI parameters.  
We find first that there are  generically four large-volume points (though special cases have three) corresponding to nonlinear sigma models on incidence correspondences, and we verify that the vanishing of the first Chern class of $X_A$ is equivalent to the condition that the axial $U(1)$ R-symmetry of the GLSM be non-anomalous, as expected. 
Using the methods developed in \cite{Hori:2006dk}, we determine the ``singular locus'' in parameter space --- this is a set of divisors along which the superconformal field theory becomes singular due to the emergence of non-compact Coulomb or mixed Coulomb--Higgs branches.  The analysis is more complicated than the abelian GLSM since there are classical branches where nonabelian gauge bosons become massless. We provide evidence supporting the hypothesis that these branches are lifted quantum mechanically, as in \cite{Hori:2006dk}.

We apply our methods to analyze some explicit examples of determinantal Calabi--Yau manifolds that have appeared in the literature \cite{Horrocks1973, Schoen, Bertin0701, Kapustka0802,gross2001calabi}, see also \cite{Hosono:2011vd}. In particular, our analysis agrees exactly with \cite{Hosono:2011vd}, where Hosono and Takagi (en route to another example) studied the quantum moduli space of a resolved determinantal quintic in $\P^4$ using mirror symmetry.  In \cite{Bertin0701}, Bertin constructed examples of codimension four Calabi--Yau varieties in $\P^7$, two of which are determinantal and can be analyzed using this GLSM; we have computed the singular loci in these examples, providing a prediction that can be checked against future studies of these manifolds using mirror symmetry.  As a byproduct we also determine various topological invariants of these examples, such as intersection numbers.

\section{Determinantal Varieties}
\label{sec:det-var}
\def\Xk{Z(A,k)}
\def\Xki{\widetilde Z(A,k)}
\def\Vpt{\phi}
\def\Gpt{x}
In this section, we collect and discuss some mathematical properties of determinantal varieties.   A discussion of many of the relevant concepts can be found in \cite{Harris1995,Fulton1997}.  On a first reading, readers may consider skipping the more technical details of this section.

To define determinantal varieties, we begin with a compact algebraic variety $V$ of (complex) dimension $D$ with two vector bundles $\cE$ and $\cF$ of rank $n$ and $m$ and a linear map
\begin{equation} \label{eq:LinA}
  A: \ \cE \rightarrow \cF \ .
\end{equation}
$A$ is a (generic) global holomorphic section of the rank $nm$-bundle $\Hom(\cE,\cF)\cong \cE^* \otimes \cF$, which we require to be generated by its global holomorphic sections.\footnote{This ensures that the (generic) linear map $A$ has maximal rank at a generic point of $V$. One can study determinantal varieties defined by a linear map between bundles which do not satisfy this property, but the resulting variety may fail to be smooth, even when $A$ is generic.  This may be related to the difficulties found by Kanazawa \cite{kanazawa2010pfaffian} in locating a crepant resolution for certain ``Pfaffian-mirror'' Calabi--Yau varieties.}  Locally, we can view the linear map $A$ as a $m\times n$-matrix of holomorphic sections. Then the determinantal variety $Z(A,k)$ of rank less than or equal to $k$, where $0\le k<\min(n,m)$, is defined as
\begin{equation} \label{eq:detX}
   \Xk \,=\,  \{ \ \Vpt \in V \ | \ \rank\,A(\Vpt) \leq k\ \}\  .
\end{equation}
$\Xk$ is associated to the ideal $I(\Xk)$ generated by all the $(k+1)\times (k+1)$ (determinant) minors of $A$. For $k>0$, the set of minors fails to intersect transversely, reflecting the fact that the minors satisfy non-trivial relations when viewed as elements of the ideal $I(Z(A,k))$.  As a consequence, $Z(A,k)$ is not a complete intersection in general: the number of $(k+1)\times (k+1)$ minors typically exceeds the codimension in $V$, which is given by
\begin{equation} \label{eq:codim}
  \codim\,\Xk \,=\, (m-k)(n-k) \ .
\end{equation}

Singularities on the subvariety $Z(A,k)$ arise at points where the rank of the Jacobian matrix $J(Z(A,k))$ drops below the codimension $(m-k)(n-k)$ of $Z(A,k)$ in $V$. In a patch of $V$ with local coordinates $(z_1,\ldots,z_D)$, the Jacobian matrix is
\begin{equation}
  J(Z(A,k))(z) \,=\, \left[ \frac{\partial \det m_s(z)}{\partial z_t} \right]_{s=1,\ldots, N\!;\ t=1,\ldots,D}
  \,=\, \left[ \tr \left({\rm adj}(m_s(z)) \frac{\partial m_s(z)}{\partial z_t} \right)\right]_{s=1,\ldots, N\!;\ t=1,\ldots,D} \ ,
\end{equation}
where $m_1(z),\ldots,m_N(z)$ denote the $(k+1)\times (k+1)$-submatrices of the matrix $A$, $N={m \choose k+1} {n \choose k+1}$. Also, ${\rm adj}(m_s)$ denotes the adjugate matrix of $m_s$, whose entries are the maximal minors of the matrices $m_s$ (equivalently, the $k\times k$-minors of the matrix $A$). We see immediately that $Z(A,k)$ is singular along $Z(A,k-1)$ (when it is non-empty) since the Jacobian matrix vanishes identically there:
\begin{equation} \label{eq:Sing}
  Z(A,k-1) \, \subseteq \, {\rm Sing}(\Xk) \, \subset \, \Xk \ .
\end{equation}
Since we have assumed that $\Hom(\cE,\cF)$ is generated by its global sections, all singularities of $Z(A,k)$ will arise from $Z(A,k-1)$ or from singularities induced from the ambient variety $V$.

There is a natural birational desingularization of the determinantal variety $Z(A,k)$ through the incidence correspondence \cite{Harris1995}
\begin{equation} \label{eq:Xkione}
  X_A := \widetilde{Z}(A,k) :=\, \left\{\ (\Vpt,\Gpt)\in V_{\cE,n-k} \ \big| \ A(\Vpt)\Gpt=0\ \right\} \ .
\end{equation}
Here $V_{\cE,n-k}$ denotes the fibration
\begin{equation} \label{eq:defV}
  G(\ell,\cE) \longrightarrow V_{\cE,\ell}  \stackrel{\pi}{\longrightarrow} V \ ,
\end{equation}
of (complex) Grassmannians of $\ell$-planes with respect to the $n$-dimensional fibers of the rank $n$ vector bundle $\cE$, and the pair $\Vpt$ and $\Gpt$ refer to coordinates on the base and the Grassmannian fiber, respectively.\footnote{This notation is a bit schematic as $\Gpt$ denotes a point in the Grassmannian fiber. The relation $A(\phi)\Gpt=0$ in~\eqref{eq:Xkione} requires any vector in the Grassmannian plane $\Gpt$ to be in the kernel of $A(\phi)$.}  
In this construction, $\phi \in V \backslash \Xk$ will never correspond to a point in $X_A$ since, by definition, at such points $\rank (A)>k$. At a smooth point $\Vpt \in Z(A,k)$, $A$ has rank $k$, hence there is a unique $(n-k)$-plane $x$ in the kernel of $A$ that gives rise to a unique point $(\Vpt,\Gpt) \in X_A$.  When the rank of $A$ drops to $k-l$ for $0< l \leq k$, the kernel of $A$ will have dimension $(l+n-k)$, so $x$ will correspond to the space of $(n-k)$-planes within this $(l+n-k)$-dimensional space; this is isomorphic to $G(n-k,l+n-k)$ and resolves the singularity associated with $Z(A,k-l)$. Thus, the incidence correspondence constructs a resolved variety $X_A$ that is birational to $Z(A,k)$.  For a more concrete discussion, including examples, see sections \ref{sec:linear-det-var} and \ref{sec:examples}.

Following \cite{Tjotta1999}, there is a dual construction of a desingularized variety $\hat{X}_A$ involving the Grassmannian fibers $G(k,\cE^*)$ dual to $G(n-k,\cE)$. To this end, consider the rank $k$ bundle $\cU$ and the rank $n-k$ bundle $\cQ$ over the variety $V_{\cE^*,k}$, which restrict to the universal subbundle and the universal quotient bundle over each Grassmannian fiber $G(k,\cE^*)$, respectively.\footnote{The universal subbundle $\cU$ is a rank $k$ bundle over the Grassmannian $G(k,n)$ and its fiber $\cU_{\tx}$ over a point $\tx$ in $G(k,n)$ is the $k$-plane representing the point $\tx$. $\cQ$ is the rank $n-k$ universal quotient bundle $\mathbb C^n/\cU$ of $G(k,n)$.\label{foot:universal}} Over $V_{\cE^*,k}$\,,  \,$\cU$ is a subbundle of $\pi^*\cE^*$ and we  can define the bundle
\begin{equation}
    \cX \,:=\, (\pi^*\cE^*/\cU)\otimes\pi^*\cF 
    \, \cong \, \cQ \otimes \pi^*\cF \, \cong \,
    \Hom\left(  \cQ^*,\pi^*\cF \right) \ ,
 \end{equation}
where the quotient $\pi^*\cE^*/\cU$ is identified with the universal quotient bundle $\cQ$.  By construction, the bundle $\cX$ is of rank $(n-k)m$ with a global holomorphic (matrix) section $\tilde A$ induced from the (matrix) section $A$ in \eqref{eq:LinA} as follows.  Tensor the canonical short exact sequence of the Grassmannian $G(k,\cE^*)$ with $\pi^*\cF$, yielding the short exact sequence $0\rightarrow \cU \otimes\pi^*\cF{\stackrel \iota \hookrightarrow} \pi^*\cE^* \otimes \pi^*\cF {\stackrel p \rightarrow} \cQ \otimes \pi^*\cF \rightarrow 0$.  Then the section $A: V\rightarrow \cE^*\otimes\cF $ naturally pulls back to a section $\pi^*A: V_{\cE^*,k} \rightarrow \pi^*\cE^* \otimes \pi^*\cF$ that, in turn, induces a section
\be
 \tilde A\! :~~ V_{\cE^*,k} \rightarrow \cQ \otimes \pi^*\cF\, , ~~ (\phi,\tx)\mapsto \tilde A(\phi,\tx):=p\circ \pi^*A(\phi,\tx) \ .
 \ee

At a point $\Vpt \in V$ with $\rank\,A(\Vpt)>k$, the section $\tilde A(\Vpt,\tx)$ is non-zero for all $\tx \in G(k,\cE^*_\Vpt)$; at a point $\Vpt$ with $\rank\,A(\Vpt)=k$, $\tilde A(\phi,\tx)$ is zero for precisely one point $\tx\in G(k,\cE^*_\Vpt)$; at a point $\Vpt \in V$ with $\rank\,A(\Vpt)=k-l<k$, there is a space $G(l,l+n-k)$ of $l$-planes within the $(l+n-k)$-dimensional kernel of $A$ along which $\tilde A$ vanishes. Thus, the desingularized variety
\begin{equation} \label{eq:Xkitwo}
  \hat{X}_A \,=\, 
  \{ \ (\Vpt,\tx)\in V_{\cE^*,k}\  |\   \tilde A(\Vpt,\tx)=0 \ \} \ ,
\end{equation}
is of dimension $\dim V + \dim G(k,n) - \rank\,\cX= D - (n-k)(m-k) = \dim \Xk$ and is again a birational resolution of the variety $\Xk$. Moreover, since the ambient varieties $V_{\cE^*,k}$ and $V_{\cE,n-k}$ are naturally dual to each other, the incidence correspondences \eqref{eq:Xkione} and \eqref{eq:Xkitwo} yield isomorphic desingularized varieties:
\be
\hat{X}_A \cong X_A \, .
\ee
Since they are isomorphic, we may on occasion use the two symbols interchangeably.  We should emphasize that the incidence correspondence \eqref{eq:Xkitwo} realizes the desingularized determinantal variety $\widetilde Z(A,k)$ as a \emph{complete intersection} in $V_{\cE^*,k}$.

Note that there are two more incidence correspondences associated to dual linear forms of $\Hom(\cF^*,\cE^*)$ of the transposed section $A^T$ .  The corresponding desingularized varieties are constructed analogously to \eqref{eq:Xkione} and \eqref{eq:Xkitwo} and arise as subvarieties of $V_{\cF,m-k}$ and $V_{\cF^*,k}$, respectively.

Resolving the variety $\Xk$ with the incidence correspondence \eqref{eq:Xkitwo} allows us to directly compute topological invariants of the resolved variety $\hat{X}_A$. Namely, since $\cX$ is the normal bundle of the (smooth) subvariety $\hat{X}_A \subset V_{\cE^*,k}$, the total Chern class of $\hat{X}_A$ is given by
\begin{equation}
   c(\hat{X}_A)\,=\, 1+\sum_{d=1}^{\dim \hat{X}_A} c_d(\hat{X}_A) \,=\, \frac{c(V_{\cE^*,k})}{c(\cX)} \ .
\end{equation}
In particular, the first Chern class of $\hat{X}_A$ becomes 
\begin{equation} \label{eq:vanChern}
\begin{aligned}
  c_1(\hat{X}_A) \,&=\, c_1(V_{\cE^*,k}) - c_1(\cX) \\
   \,&=\, \pi^*c_1(V) -  (m-k)\,\pi^*c_1(\cE^*)- (n-k)\,\pi^*c_1(\cF)-(n-m)c_1(\cU) \ ,
\end{aligned}
\end{equation}
with $c_1(\cX)=m\,\pi^*c_1(\cE^*)+(n-k)\,\pi^*c_1(\cF)-m\,c_1(\cU)$ and $c_1(V_{\cE^*,k})=\pi^*c_1(V)+k\,\pi^*c_1(\cE^*)-n\,c_1(\cU)$. The Euler characteristic $\chi(\hat{X}_A)$ can be evaluated by use of the relation:
\begin{equation} \label{eq:Euler}
   \chi(\hat{X}_A) \,=\, \int_{\hat{X}_A} c_{top}(\hat{X}_A)\\
   \,=\, \int_{V_{\cE^*,k}} c_{top}(\cX) \wedge c_{top}(\hat{X}_A) \ .
\end{equation}
Furthermore, the intersection numbers $I(\gamma_{k_1},\gamma_{k_2},\ldots,\gamma_{k_s})$ for cohomology classes $\gamma_k$ in $H^{\rm ev}(\hat{X}_A,\Z)$, which are induced from the ambient cohomology group $H^{\rm ev}(V_{\cE^*,k},\Z)$, are computed by
\begin{equation} \label{eq:Inter}
  I(\gamma_{k_1},\ldots,\gamma_{k_s})
  \,=\, \int_{\hat{X}_A} \gamma_{k_1} \wedge \ldots \wedge \gamma_{k_s}
  \,=\, \int_{V_{\cE^*,k}} c_{top}(\cX) \wedge \gamma_{k_1} \wedge \ldots \wedge \gamma_{k_s} \ .
\end{equation}

Let us remark that these intersection calculations become more feasible when the ambient variety $V_{\cE^*,k}$ is a trivial fibration and, thus, a product variety $V \times G(k,n)$.  This happens when the rank $n$ vector bundle $\cE$ is equivalent to a tensor product
\begin{equation} \label{eq:simpleE}
  \cE \cong \cL \otimes \cO_V^{\oplus n} \ ,
\end{equation}
where $\cO_V$ is the structure sheaf of $V$ and $\cL$ is a line bundle. Then the bundle $\cX$ becomes $\Hom(\cQ^*,\cL^*\otimes\cF)$ and the two incidence correspondences \eqref{eq:Xkione} and \eqref{eq:Xkitwo} simplify respectively to
\begin{equation} \label{eq:wXkthree}
  X_A \,\longrightarrow\,  \big\{\ (\Vpt,\Gpt)\in V\times G(n-k,n) \ \big| \ A(\Vpt)\Gpt=0\ \big\} \, 
\end{equation}
and
\begin{equation} \label{eq:wXkfour}
  \hat{X}_A \,\longrightarrow\,  \big\{\ (\Vpt,\tx)\in V\times G(k,n) \ \big| \ \tilde A(\Vpt,\tx)=0\ \big\} \, .
\end{equation}
For many examples discussed in this work, \eqref{eq:simpleE} holds and the simplified incidence correspondences become applicable.  This also includes linear determinantal varieties in projective spaces, furnishing an important class of examples \cite{Harris1995}.

In the following we concentrate on (complex) three-dimensional Calabi--Yau determinantal varieties, so $\dim X_A= D-(n-k)(m-k)=3$ and $c_1=0$.  For three-dimensional varieties, the dimension of the singular locus \eqref{eq:Sing} is $2k-(n+m-2)$ with $k<\min(n,m)$.  Hence, for threefolds the generic singular locus can only be non-empty if $n-1=m-1=k$, in which case it would be zero-dimensional.  Otherwise, there are no loci of reduced rank for such three-dimensional determinantal varieties with generic $A$.

Nevertheless, even in situations with no singular loci the incidence correspondence \eqref{eq:Xkitwo} proves useful in calculating topological invariants of determinantal varieties.  Recall that the diffeomorphism class of simply connected Calabi--Yau threefolds $X$ (with $H^3(X,\Z)$ torsion free) is determined by: (\emph{i}) the cubic intersection form $H^2(X,\Z)\times H^2(X,\Z)\times H^2(X,\Z) \longrightarrow \Z$; (\emph{ii}) the linear form $H^2(X,\Z){\stackrel  {c_2} \longrightarrow} \Z$ given by the cup product with the second Chern class $c_2(X)$; and, (\emph{iii}) the middle-dimensional cohomology group $H^3(X,\Z)$ \cite{Wall1966}. Therefore, getting a handle on intersection numbers of determinantal Calabi--Yau threefolds can help us to identify its topological type.

\section{Nonabelian GLSMs for Determinantal Varieties}

In Section \ref{sec:overview}, we summarized the setup for the PAX and PAXY models.  In Table \ref{tab:PAX-PAXY}, we list the charges of the fields for each model. In the following subsections, we explain each in greater detail.  For reasons given in Section \ref{sec:generalize}, we will focus on determinantal varieties defined by \emph{square} $n\times n$ matrices $A$, unless stated otherwise.
\begin{table}[t]
\hspace{-1.2cm}
{\bf PAX}:
\hspace{0.1cm}
\begin{tabular}{|c|c|c|}
\hline
	& $U(\kd)$ 	& $U(1)_\ell$ \\
\hline
$X_i$	&  $\overline{\mathbf{\kd}}$ 	& $-q_i^\ell$ \\
$P_i$	& $\mathbf{\kd}$	& $-\mathfrak{q}_i^\ell$ \\
$\Phi_a$	& $\mathbf{1}$		&	$Q_a^\ell$ \\	
\hline
\end{tabular}
\hspace{1.25cm}
{\bf PAXY}:
\hspace{0.1cm}
\begin{tabular}{|c|c|c|}
\hline
	& $U(k)$ 	& $U(1)_\ell$ \\
\hline
$\widetilde{X}_i$	&  $\mathbf{k}$ 	& $q_i^\ell$ \\
$\widetilde{Y}_i$	& $\overline{\mathbf{k}}$	& $\mathfrak{q}_i^\ell$ \\
$\tP_{ij}$ & $\mathbf{1}$		& $-(q_i^\ell+\mathfrak{q}_j^\ell)$ \\
$\Phi_a$	& $\mathbf{1}$		&	$Q_a^\ell$ \\	
\hline
\end{tabular}
\centering
\caption{On the left, chiral superfields in the PAX model and their representations under the gauge group $U(\kd) \times U(1)^s$.  On the right, the same for the PAXY model.  In both cases, $\mathbf{\Delta}$ and $\overline{\mathbf{\Delta}}$ denote the fundamental and anti-fundamental representations, while $i,j \in \{1,\ldots,n\}$, $a \in \{ 1,\ldots ,D+s\}$, and $\ell\in\{1,\ldots,s\}$.} \label{tab:PAX-PAXY}
\end{table}

\subsection{Axial Anomaly and Central Charge}
\label{sec:anomalies}

We are interested in GLSMs that flow to a $(2,2)$ superconformal field theory describing string propagation on a determinantal Calabi--Yau. 
Since the $(2,2)$ superconformal algebra contains an axial R-symmetry, it is convenient to require that the $U(1)$ axial R-symmetry in the GLSM be anomaly free; in the nonlinear sigma model, the axial R-symmetry is anomaly-free only when the first Chern class of the target space vanishes, so the target space must be Calabi--Yau (if it is K\"ahler).  
In two dimensions, such an anomaly will arise from a one-loop diagram with one insertion of a gauge current and one insertion of the axial current. Here we will see how the anomalies between our two dual formulations coincide. In the PAXY model with matter content summarized in Table \ref{tab:PAX-PAXY}, the axial anomaly from the $U(1)^s$ gauge groups is given by
\be
\label{eqn:ambient-axial-anomaly}
\sum_{a=1}^{D+s} Q_a^\ell - \sum_{i,j=1}^{n} \big(q_i^\ell+\mathfrak{q}_j^\ell\big) + k \sum_{i=1}^n \big( q_i^\ell + \mathfrak{q}_i^\ell \big) \ = \
\sum_{a=1}^{D+s} Q_a^\ell - (n-k) \sum_{i=1}^n \big( q_i^\ell + \mathfrak{q}_i^\ell \big)
= 0  \ ,
\ee
for all $\ell=1,\ldots,s$.  Note that the middle expression is also equivalent to the anomaly in the PAX model. If we denote the charge under $U(1)_{\ell}$ by $\mathrm{deg}_{\ell}$, gauge invariance of the superpotential imposes the condition $\textrm{deg}_{\ell}\!\left(A(\phi)_{ji}\right) = \mathfrak{q}_j^\ell+q_i^\ell$, so $\mathfrak{q}_j^\ell+q_i^\ell$ must be expressible as a non-negative, integral linear combination of $Q_a^\ell$.  This fact allows us to equivalently write the anomaly cancellation condition as
\be
\sum_{a=1}^{D+s} Q_a^\ell = (n-k)\, \deg_{\ell}\Big( \!\!\det\!\big(A(\phi)\big)\!\Big) \, .
\ee
Since we have restricted to square matrices $A(\phi)$, the anomaly also cancels for $U(1) = \det\big(U(k)\big)$ in the PAXY model and for $U(1) =\det\big(U(\kd)\big)$ in the PAX model since there are an equal number of fundamental and anti-fundamental chiral superfields charged under each $U(1)$.  Therefore, the conditions for anomaly cancellation are identical in the PAX and PAXY models.

Note that this anomaly cancellation condition is in agreement with the geometric Calabi--Yau condition \eqref{eq:vanChern} of a vanishing first Chern class.  The first term in the anomaly condition \eqref{eqn:ambient-axial-anomaly} corresponds to the first Chern class $c_1(V)$ of the toric variety $V$, whereas the charges $q_i^\ell$ and $\mathfrak{q}_j^\ell$ are associated to the first Chern classes of line bundles whose direct sum gives rise to the rank $n$ and rank $m$ vector bundles $\cE^*$ and $\cF$, respectively. Thus, the sums $\sum q_i^\ell$ and $\sum \mathfrak{q}_j^\ell$ map to the first Chern classes $c_1(\cE^*)$ and $c_1(\cF)$. Finally, the cancellation of the $\det\big(U(k)\big)$ or $\det\big(U(\kd)\big)$ anomaly reflects the fact that the summand $c_1(\cU)$ in \eqref{eq:vanChern} drops out for $n=m$.

The central charge of the IR superconformal field theory is computed from the $U(1)$ axial-vector current two-point function, where we compute the one-loop contribution to the OPE of the axial current with the vector current. Both axial and vector charges can always be shifted by a linear combination of the gauge charges without affecting the anomalies or central charge so, as long as anomaly cancellation is satisfied, we basically only need to know the number of fields.  In both models, we can take the vector charges of fields to be an arbitrary linear combination of gauge charges, except for the $P$ or $\tP$ fields whose vector charges we shift by $2$ to cancel the charge of the measure over half of superspace.  Noting also that the gauginos are charged under the axial and vector symmetries (ensuring that the twisted superpotential has appropriate charges), we find the central charge of the PAX model to be 
\be
\frac{c}{3} =(D+s) +\kd n-\kd n -\kd^{2} - s = D - (n-k)^2 \, ,
\ee
which is precisely what we want since the rank $k$ condition locally imposes $(n-k)^2$ constraints in the $D$-dimensional ambient variety $V$. Applying the same logic to the PAXY model, we find
\be
\frac{c}{3} = (D+s) - n^2 + 2kn - s - k^2 = D - (n-k)^2 \, ,
\ee
so both central charges coincide with what we expect.

\subsection{The PAX model}
\label{sec:PAX}

The Lagrangian for the GLSM with matter content given by Table \ref{tab:PAX-PAXY} is
\begin{eqnarray}
L = \int d^4\theta\ K + \left(\int d^2\theta \ W + \int d^2\tilde{\theta}\ \tW + c.c.\right)\, ,
\end{eqnarray}
where we will follow superspace and superfield conventions of \cite{Witten:1993yc, Witten:1993xi}.  The K\"ahler potential $K$ is
\be
K = \frac{1}{4} \sum_i \left(X_i e^{-2V_{U(\kd)} - 2 q_i^\ell V_{U(1)_\ell}} X^{\dag}_i + P^{\dag}_i e^{2V_{U(\kd)}-2\mathfrak{q}_i^\ell V_{U(1)_\ell}} P_i \right)   + \frac{1}{4} \sum_a \Phib_a e^{2Q_a^\ell V_{U(1)_\ell}}\Phi_a\, ,
\ee
where the fields $P_i$ and $X_i^T$ are in the $\mathbf{\kd}$ and $\overline{\mathbf{\kd}}$ of $U(\kd)$, respectively. $V_{U(\kd)}$ and $V_{U(1)_\ell}$ are the vector superfields associated with the $U(\kd)$ and $U(1)^s$ factors in the gauge group, and $V_{U(\kd)}$ is in the fundamental representation (we have made use of the fact that for representation matrices of $\mathfrak{u}(\kd)$, $T_{\overline{\mathbf{\kd}}} = -(T_{\mathbf{\kd}})^*$).  The superpotential is given by (\ref{eqn:PAX}), while the twisted superpotential is
\begin{equation}
2\sqrt{2}\,\tW =  it_{U(\kd)} \, \tr \Sig_{U(\kd)} + i\sum_{\ell=1}^s  t_\ell \, \Sig_{U(1)_\ell}\ ,
\end{equation}
where $\Sig_{U(\kd)}$ and $\Sig_{U(1)_\ell}$ are the twisted chiral field strengths of the vector multiplets.
The parameters $t$ are the usual complexifications of the FI parameters to include theta angles
\begin{equation}
t = i r + \frac{\theta}{2\pi} \,.
\end{equation}
The theta angles are $2\pi$ periodic when the integer charges of the elementary fields have no divisor greater than one, which we will take to be the case. When the GLSM flows to a nonlinear sigma model, the complexified FI parameters correspond to the complexified K\"ahler parameters of the target space, while the superpotential determines the complex structure.  When the target space is a Calabi--Yau, the GLSM flows to a $(2,2)$ superconformal field theory in the IR and the complexified FI parameters correspond to exactly marginal deformations of the superconformal field theory. 
The D-terms and F-terms above arise from integrating out the auxiliary fields in the gauge and chiral multiplets, respectively. The D-terms give rise to the following classical relations 
\begin{eqnarray}
U(1)_\ell: & -\sum_{i}(q_{i}^{\ell}x_{i}x_{i}^{\dag}+\mathfrak{q}_{i}^{\ell}p^{\dag}_{i}p_{i})+\sum_{a}Q_{a}^{\ell}|\phi_{a}|^{2}=r_{\ell} \, ,\\
U(\kd): & pp^{\dag}-x^{\dag}x=r_{U(\kd)}\mathds{1}_{\kd} \, .
\label{eq:PAX-D}
\end{eqnarray}
The F-terms, which do not receive quantum corrections, read
\begin{eqnarray}
A(\phi)\, x=0 \, ,\qquad p\, A(\phi)=0 \, , \qquad \mathrm{tr}\left(p\,\partial_{a}A(\phi)\,x\right)=0 \, .  \label{eq:PAX-F}
\end{eqnarray}

\subsubsection*{The Geometric Phase}
Under a smoothness assumption, we can always find a geometric phase that describes the determinantal Calabi--Yau of interest, or a resolution thereof.  In this phase,  $r_{U(\kd)} < 0$ while the $r_\ell$ must lie in the interior of the K\"ahler cone of $V$ (we will justify in the next paragraph that the $r_\ell$ can be so chosen).  The first condition combines with the D-term \eqref{eq:PAX-D} to guarantee that $\rank(x)=\kd$ or, equivalently, $\det(x^T x) \neq 0$.  The fields $(\phi,x,p)$ then form the total space of a vector bundle: $(\phi,x)$ are coordinates on the base space $V_{\cE,\kd}$ (which is locally a product of $V$ and $G(\kd,n)$)
while the $p$ fields are fibered over $V_{\cE,\kd}$ as dictated by their representation in Table \ref{tab:PAX-PAXY}. The F-term condition $A(\phi)\, x=0$ implies that $\rank(A)\leq k$ with $x \in \ker(A)$, so we arrive at the desired incidence correspondence:
\be \label{eq:inPAX}
  X_A := \,\big\{\ (\phi,x)\in V_{\cE,\kd}\ \big|\ A(\phi) x=0\ \big\}  \, .
\ee
At smooth points, $X_A$ is defined by the rank of its normal bundle being equal to its codimension in $V_{\cE,\kd}$, namely $n(n-k)$. Defining $E_{i\alpha} := A_{ij}(\phi) x_{j\alpha}$, the normal bundle is then spanned by the $n\,\kd=n(n-k)$ vectors, $\partial_{(\phi,x)}(E_{i \alpha}) |_{X_A}$, and so they must be linearly independent if $X_A$ is smooth. The remaining F-term conditions, both of which involve $p$, can then be rewritten as
\be
p_{\alpha i} \frac{\partial E_{i \alpha}}{\partial(\phi_{a},x_{j\beta})}=0\, .
\ee
Therefore, smoothness of $X_A$ implies $p=0$. The solution of the D-terms and F-terms for the range of FI parameters specified above is precisely the incidence correspondence \eqref{eq:inPAX}.  When $Z(A,k) \subset V$ is smooth, then $X_A$ is isomorphic to $Z(A,k)$ yielding a smooth determinantal variety.  Otherwise, the singularities of $Z(A,k)$ arising from a rank degeneration are resolved by the incidence correspondence $X_A$.

Now we return to the claim that $r_\ell$ can be suitably chosen so that $\phi_a$ will yield a compact variety $V = \big( \mathbb{C}^{D+s} \backslash F\big) /\big(\mathbb{C}^*\big)^s$.  Assuming $A(\phi)$ can be chosen so that $X_A$ is smooth for $\phi_a \notin F$, then $p=0$.  However, when $\phi_a \in F$, it may not be the case that $p$ is forced to vanish (for example, when $\phi_a=0$ for all $a$).  To avoid this possibility, we must check that $r_\ell$ can be chosen so that $\phi_a \notin F$.  To that end, define $q_-^\ell := \min_i\{q_i^\ell\}$ and $\mathfrak{q}_-^\ell := \min_i\{\mathfrak{q}_i^\ell\}$.  Then the D-terms imply 
\be
\sum_a Q_a^\ell |\phi_a|^2  \, \geq \, r_\ell + q_-^\ell \tr(x x^\dag) + \mathfrak{q}_-^\ell \tr(p^\dag p) \, = \, r_\ell + \kd q_-^\ell |r_{U(\kd)}| + (q_-^\ell + \mathfrak{q}_-^\ell) \tr(p^\dag p) \, .
\ee
For ease of exposition, we will take $Q_a^\ell \geq 0$ (though the qualitative results should be the same whenever $V$ is compact). Since $(q_-^\ell + \mathfrak{q}_-^\ell)$ is the charge of one of the entries of $A(\phi)$, it too must be nonnegative since it must be a nonnegative integral linear combination of the $Q_a^\ell$.  We therefore find 
\be
\sum_a Q_a^\ell |\phi_a|^2 \, \geq \, r_\ell + \kd q_-^\ell | r_{U(\kd)}| \, ,
\ee
so we can indeed tune the FI parameters to ensure that $\phi_a \notin F$, forcing $p$ to vanish in this phase and yielding the incidence correspondence $X_A$, as claimed.

\subsection{The PAXY model}
\label{sec:PAXY}

The PAXY model, with gauge group $U(1)^s \times U(k)$, matter content as in Table \ref{tab:PAX-PAXY}, and K\"ahler potential similar to the PAX model, has a superpotential given by \eqref{eqn:nonabel-spot} and a twisted superpotential
\begin{equation}
2\sqrt{2}\,\tW =  it_{U(k)} \, \tr \Sig_{U(k)} + i{\textstyle \sum_{\ell=1}^k}t_\ell \, \Sig_{U(1)_\ell}\ .
\end{equation}
The D-terms give rise to the following classical relations:
\begin{eqnarray}
U(1)_\ell: & \sum_i \big( q_i^\ell \ \tx_i^\dag \tx_i + \mathfrak{q}_i^\ell \ \ty_i \ty_i^\dag \big)  + \sum_a Q_a^\ell |\phi_a|^2 - \sum_{i,j} \big(q^\ell_i + \mathfrak{q}^\ell_j \big) |\tp_{ij}|^2 = r_{\ell} \, , \\
U(k): & \tx\tx^\dag - \ty^\dag \ty = r_{U(k)} \mathds{1}_k \, .
\end{eqnarray}
The F-terms, which do not receive quantum corrections, read
\begin{eqnarray}
\tp \ty = 0\, , \qquad \tx \tp = 0, \qquad \tr\left(\tp \, \partial_a A(\phi)\right) = 0\, , \qquad \ty\tx - A(\phi) = 0\, . 
\end{eqnarray}

\subsubsection*{The Geometric Phase}
Assuming that $r_{U(k)} > 0$, the $U(k)$ D-term forces $\rank(\tx)=k$.  Similar to the PAX model, one can verify that the $r_{\ell}$ can be chosen to lie in the interior of the K\"ahler cone of $V$ to ensure that $\phi_a\notin F$, thus $(\phi,\tx)$ are homogeneous coordinates on $V_{\cE^*,k}$\,, which is locally a product of $V$ and $G(k,n)$ (note that it is $\cE^*$ here instead of $\cE$ since $\tx$ has the opposite $U(1)^s$ charges from $x$).  The $\ty$ are then the fiber coordinates on the rank $kn$ bundle $\cU \otimes (\oplus_i \pi^*\cO_V(\mathfrak{q}^\ell_i))$ over $V_{\cE^*,k}$, where $\cU$ projects down to the universal sub-bundle on $G(k,n)$.  The F-terms for $P$ enforce
\be
\label{eq:PAXY-F}
A(\phi) = \ty \tx \, ,
\ee
which implies that $\rank(A(\phi)) = \rank(\ty) \leq k$.  (In fact, since $\tx$ has maximal rank we can uniquely solve for $\ty$ as a function of $A(\phi)$ and $\tx$, and we will do so shortly.)  This defines an intersection of $n^2$ hypersurfaces in the total space of $\cU \otimes (\oplus_i \pi^*\cO_V(\mathfrak{q}^\ell_i))$.  For generically chosen $A(\phi)$, this intersection will be smooth which will ensure that $p$ vanishes through the remaining F-terms (as in the PAX model).  We will explain next how this forms a resolution of $Z(A,k)$, but first note that this fact implies that the dimension should be $(D-(n-k)^2)$ --- the same as the dimension of $Z(A,k)$ away from singular points --- which together with the smoothness condition implies that this must be a \emph{complete intersection} in the total space of $\cU \otimes (\oplus_i \pi^*\cO_V(\mathfrak{q}^\ell_i))$ since its codimension $n^2$ equals the number of generators of its ideal sheaf.\footnote{In fact, this observation applies more broadly since the F-terms are of the same form for all rectangular matrices: any rectangular determinantal variety defined in an ambient toric variety with $\cE^*$ and $\cF^*$ given by sums of line bundles, is birational to a complete intersection of $mn$ equations in the $(D+km+kn-k^2)$-dimensional total space of the bundle $\cU \otimes (\oplus_i \pi^*\cO_V(\mathfrak{q}^\ell_i))$ over $V_{\cE^*,k}$.}

Whenever \eqref{eq:PAXY-F} holds, $\ker\,\tx\subseteq\ker A$.  Since $\tx$ is a rank $k$, $k\times n$ matrix, we can think of $\tx$ as being defined by its $(n-k)$-dimensional kernel (up to $GL(k)$ transformations).  When $\rank(A(\phi)) = k$, $\dim\ker A = (n-k)$, so $\tx$ is defined by the set of $(n-k)$-planes in the $(n-k)$-dimensional kernel of $A$; this is simply a point which means that $\phi\in Z(A,k)\backslash Z(A,k-1)$ defines a unique point $(\phi,\tx)\in V_{\cE^*,k}$.  When $\rank(A)=k-l < k$, the kernel of $\tx$ defines an $(n-k)$-plane in the $(n-k+l)$-dimensional kernel of $A$, which thus defines a space isomorphic to $G(n-k,n-k+l)$, resolving the singularity along $Z(A,k-l)$ in the same way as the PAX model.

Since $\ty$ is uniquely determined by $A(\phi)$ and $\tx$ through \eqref{eq:PAXY-F}, we can slightly rephrase this discussion by solving for $\ty$ and defining the variety to be embedded in $V_{\cE^*,k}$.  The fact that $\tx$ has maximal rank implies that $\tx\tx^T$ is invertible, so
\be
\label{eq:PAXY-sans-Y}
\ty = A(\phi) \tx^T ( \tx\tx^T)^{-1} \, , \qquad  A(\phi) \big( \mathbf{1}_n - \tx^T(\tx\tx^T)^{-1}\tx \big) = 0 \, .
\ee
As a complex manifold, then, the solution to the F-terms is isomorphic to\footnote{Metrically, these still depend on the details of the embedding in $\cU \otimes (\oplus_i \pi^*\cO_V(\mathfrak{q}^\ell_i))$, \ie on the kinetic terms and D-terms for $\ty$ as well as $\tx$ and $\phi$, so this should just be taken as a statement about complex structures.}
\be
\label{eq:Vek-Incidence}
\hat{X}_A := \big\{ \ (\phi,\tx)\in V_{\cE^*,k} \ \big| \ A(\phi) \big( \mathbf{1}_n - \tx^T(\tx\tx^T)^{-1}\tx \big) = 0 \ \big\} \,  \subset V_{\cE^*,k} \, .
\ee
In fact, $(\mathbf{1}_n - \tx^T(\tx\tx^T)^{-1}\tx)$ is a projector onto $\ker\,\tx$ and thus has rank $(n-k)$.  We can therefore define a maximal rank $n\times(n-k)$ matrix $x$ by
\be
\label{eq:xxtilde}
(\mathbf{1}_n - \tx^T(\tx\tx^T)^{-1}\tx) =: x (x^Tx)^{-1} x^T \, ,
\ee
which is unique up to a $GL(n-k)$ action acting from the right on $x$.  In fact, this equation also implies that
\be
\tx x = 0 \, ,
\ee
which is precisely the statement that $x$ defines the $(n-k)$-plane orthogonal to the $k$-plane defined by $\tx$.  We therefore can write
\be
\label{eq:PAXY-is-PAX}
\hat{X}_A  \ \cong \ \big\{ \ (\phi,x) \in V_{\cE,n-k} \ \big| \ A(\phi) x = 0 \ \} = X_A \, .
\ee
We see, then, that the geometric phases of PAX and PAXY yield moduli spaces that are isomorphic as complex manifolds.

We can connect this discussion with the incidence correspondence \eqref{eq:Xkitwo} in the following way:  since $\ker\,\tx$ can be identified with $\cQ_{(\phi,\tx)}$, the projector $(\mathbf{1}_n - \tx^T(\tx\tx^T)^{-1}\tx)$ is nothing but the map $p$ that defines the section $\tilde A$ of $\cX$ in \eqref{eq:Xkitwo}.  Thus, $\tilde A$ may be viewed as representing the equivalence classes of matrices $A(\phi)(\mathbf{1}_n - \tx^T(\tx\tx^T)^{-1}\tx)$ at a given point $(\phi,\tx)\in V_{\cE^*,k}$ and therefore is zero only if \eqref{eq:PAXY-sans-Y} holds. Hence, the incidence correspondence \eqref{eq:Vek-Incidence} is an equivalent formulation of \eqref{eq:Xkitwo}, and the duality between the PAX and PAXY models in the large volume phase realizes the duality between the two distinct but isomorphic incidence correspondences described in Section~\ref{sec:det-var}.

\subsection{Duality between PAX and PAXY}

We have demonstrated that the PAX and PAXY models have isomorphic vacuum moduli spaces  in geometric phases of FI parameter space. 
In this section we observe that the two models are related by a two-dimensional version of Seiberg duality \cite{Seiberg:1994pq}. These two-dimensional dualities were first discussed by Hori and Tong for the gauge group $SU(k)$ with fundamental matter, and later extended by Hori for the case of symplectic and orthogonal groups \cite{Hori:2006dk, Hori:2011pd}. The latter examples are quite interesting as the rules of \cite{Berenstein:2002fi} do not apply, essentially because gauge invariant meson fields can be constructed using fields in the fundamental representation alone. When the gauge group is $U(k)$, mesons can only be formed if we have a pair of fields transforming in the $\mathbf{k}$ and $\mathbf{\bar{k}}$ representation, and the rules of \cite{Berenstein:2002fi} can be applied in a straightforward manner. 

We introduce the mesons $M_{ij}=\sum_{\hat{\alpha}=1}^{k}\widetilde{Y}_{i\hat{\alpha}}\widetilde{X}_{\hat{\alpha} j}$ in (\ref{eqn:nonabel-spot}) as well as Lagrange multipliers $P_{\alpha i}$ and $X_{j\alpha}$ of dimensions $\kd\times n$ and $n\times\kd$, respectively, and the following superpotential
\begin{eqnarray}
\sum_{i,j=1}^{n}\widetilde{P}_{ji}(A(\Phi)_{ij}-M_{ij})+\sum_{\alpha=1}^{n-k}P_{\alpha i}M_{ij}X_{j\alpha} \, .
\end{eqnarray}
$P$ and $X$ transform as a fundamental and anti-fundamental of a $U(\kd)$ gauge symmetry.  Integrating out $P$ and $X$ forces $M$ to have rank less than or equal to $k$, yielding the original PAXY model since then we can write $M=\tY\tX$.  Alternatively, integrating out $M_{ij}$ will give us the superpotential
\begin{eqnarray}
W_{\mathrm{dual}}=\sum_{i,j=1}^{n}\sum_{\alpha=1}^{n-k}P_{\alpha i}A(\Phi)_{ij}X_{j\alpha}=\mathrm{tr}\left(PAX\right) \, ,
\end{eqnarray}
which is the same as (\ref{eqn:PAX}).  In Section \ref{sec:anomalies}, we found that the anomaly and central charge of both models matched, and in Section \ref{sec:PAXY} we found that the solutions of the F- and D-terms in geometric phases of FI parameter space are isomorphic as complex manifolds, so we conclude that the models are dual to each other.

\subsection{Generalizations}
\label{sec:generalize}

\subsubsection*{Non-square Determinantal Varieties}

In order to construct a non-square determinantal variety $\hat{X}_A$ of vanishing first Chern class, non-trivial geometric conditions must be met. As before, using the PAXY incidence correspondence \eqref{eq:Xkitwo}, we view $\hat{X}_A$ of rank $k$ as a determinantal subvariety of the ambient space $V_{\cE^*,k}$.  As opposed to the square determinantal varieties (with $n=m$), we observe that for $n\neq m$\, the class $c_1(\cU)$ of the universal quotient bundle $\cU$ of $V_{\cE^*,k}$ enters into the expression for the first Chern class of $\hat{X}_A$, \eqref{eq:vanChern}.  To obtain a Calabi--Yau threefold $\hat{X}_A$, then, the term $(m-n)c_1(\cU)$ in $c_1(\hat{X}_A)$ must either vanish or be canceled.

For non-trivial Grassmannian fibrations \eqref{eq:defV}, it is conceivable that the two-form class $c_1(\cU)$ becomes cohomologically trivial in the ambient space $V_{\cE^*,k}$ and, hence, does not contribute to $c_1(\hat{X}_A)$.  This situation, however, can only occur if the base space $V$ of $V_{\cE^*,k}$ has non-trivial $H^3(V)$.\footnote{Since $c_1(\cU)$ is a generator of the two-form cohomology of the Grassmannian fibers in $V_{\cE^*,k}$, the Leray spectral sequence implies that $c_1(\cU)$ can only become trivial in the presence of cohomologically non-trivial three-forms in the base space $V$.}  In particular, $c_1(\cU)$ must always be non-trivial when $V$ is a weighted projective space.

Alternatively, if $c_1(\cU)$ is a generator of $H^2(V_{\cE^*,k})$ then its contribution to $c_1(\hat{X}_A)$ could still be canceled against other terms in \eqref{eq:vanChern}. Such a scenario is only possible if the inclusion map $i^*: H^2(V_{\cE^*,k}) \rightarrow H^2(\hat{X}_A)$ has a non-trivial kernel.  Then the variety $\hat{X}_A$ would be Calabi--Yau if the linear combination of two forms on the right hand side of \eqref{eq:vanChern} --- viewed as a (non-trivial) linear combination of two-form classes in the ambient space $V_{\cE^*,k}$ --- resides in the kernel of the inclusion map $i^*$, hence giving rise to a cohomologically trivial first Chern class $c_1(\hat{X}_A)$. 

From the GLSM point of view, the classical analysis of D-terms and F-terms is the same for non-square matrices as it was for square matrices. However, on the quantum level the non-square cases are qualitatively different: the axial R-symmetry in the PAXY model is anomalous under $\det\!\big(U(k)\big)$ due to an unequal number of multiplets in the fundamental and anti-fundamental representations of $U(k)$.  This axial anomaly corresponds precisely to the appearance of the term $c_1(\cU)$ in the expression of the first Chern class $c_1(\hat{X}_A)$.  The same is true in the PAX model.

Varieties of codimension two which do not satisfy the ``canonical bundle restriction'' condition can be constructed as $(m+1)\times m$-determinantal varieties of rank $m-1$ \cite{Burch}, but these are never Calabi--Yau.
In fact, we are not aware of {\em any}\/ examples of  non-square determinantal Calabi--Yau varieties --- other than examples with additional symmetry properties of $A$, see e.g., \cite{Kapustka0802} --- which would allow us to study their realization in the GLSM.  However, the list of threefolds in $\mathbb{P}^5$ up to degree $18$  
 in \cite{Decker:1994} does include a few threefolds birational to Calabi--Yau manifolds, so it would interesting to study these examples with nonabelian GLSM techniques to see whether the phase structure is reproduced.

In the absence of explicit examples, we can only speculate on how a GLSM for a non-square determinantal variety could work. We can replace the gauge group $U(k)$ with $SU(k)$ and be free of the anomaly, but then we would have a noncompact direction due to the global symmetry $(\widetilde X,\widetilde Y)\rightarrow (\lambda\widetilde X,\lambda^{-1}\widetilde Y)$ for $\lambda\in\mathbb{C}^*$.  Furthermore, the presence of a global symmetry that would cause an axial anomaly if gauged means that there is no invariant definition of a central central charge for an IR theory.  Perhaps one could use the global symmetry to turn on a twisted mass for $\tX$ and $\tY$, although this would break the axial symmetry classically (since the twisted mass itself would have to be charged under an axial symmetry for invariance of the action --- think of this as an expectation value for a $\sigma$-field). This would give a mass to the noncompact direction of $\widetilde X$ and $\widetilde Y$, so if an axial symmetry happens to emerge in the infrared this would be a candidate for a non-square determinantal Calabi--Yau $\hat{X}_A$ associated to a non-trivial inclusion map $i^*: H^2(V_{\cE^*,k}) \rightarrow H^2(\hat{X}_A)$.

\subsubsection*{Symmetric and Antisymmetric Cases}

The $O(k)$ and $Sp(k)$ cases were recently discussed by Hori, and so our discussion will be brief \cite{Hori:2011pd}.  The connection here is straightforward: if $A(\phi)$ is a symmetric $n\times n$ matrix of rank $k$, it can be expressed in terms of a rank $k$, $k\times n$ matrix $\tx$ via the relation
\be
A = \tx^T \tx \, .
\ee
$\tx$ is unique up to the action of a complexified $O(k)$ symmetry, so this can be described similarly to the $U(k)$ PAXY GLSM.  The potential axial anomalies all come from the $U(1)^s$ defining the ambient variety, given by (\ref{eqn:ambient-axial-anomaly}).  Assuming anomaly-free charge assignments have been made, we simply count degrees of freedom to obtain the central charge
\be
\frac{c}{3} = \overbrace{(D+s)}^\Phi+\overbrace{kn}^{\tX} - \overbrace{{n+1\choose 2}}^{\tP} - \overbrace{{k \choose 2}}^{O(k)} - \overbrace{s}^{U(1)^s}  = D - {n-k+1\choose 2} \, ,
\ee
so the codimension in the ambient toric variety $V$ is ${n-k+1\choose 2}$, as can be confirmed by counting degrees of freedom in a symmetric $n\times n$ matrix with rank $k$.

If $A(\phi)$ is an antisymmetric, $n \times n$ matrix of rank $k=2l$ (antisymmetric matrices can only have even rank), it can be described by a rank $2l$, $2l\times n$ matrix $\tx$ via the relation
\be
A = \tx^T \left(\begin{array}{cc}  \phantom{-}\mathbf{0}_l &  \phantom{-}\mathbf{1}_l  \\  -\mathbf{1}_l & \phantom{-}\mathbf{0}_l \end{array}\right) \tx \, ,
\ee
where $\tx$ is unique up to the action of a complexified $\mathit{USp}(k)$ action.  Again, the nonabelian GLSM can be described in a similar manner as the $U(k)$ PAXY GLSM, and the axial anomaly arises solely from the $U(1)^s$ symmetries. In the absence of an axial anomaly, the central charge is given by
\be
\frac{c}{3} = (D+s)+kn-{n\choose 2} - {k+1\choose 2} - s = D - {n-k \choose 2} \, .
\ee
The codimension is thus ${n-k \choose 2}$, in agreement with degree of freedom counting for $A(\phi)$.

\subsubsection*{Other Symmetries}
In \cite{Kapustka0802}, there are examples of determinantal Calabi--Yau threefolds whose defining matrices $A(\phi)$ have assorted symmetry properties.  In Table \ref{tab:kapustka}, we reproduce many of their examples and indicate certain defining data of a corresponding PAXY GLSM.

An example that differs from what we have discussed so far is defined by the rank 2 locus of a $4\times 5$ matrix $A(\phi)$, which is obtained by deleting a row from a symmetric $5\times 5$ matrix.  At the level of a PAXY-style GLSM, one can obtain such a model with an $O(2)$ gauge group by taking $\widetilde{P}$ to be a $5\times 4$ matrix obtained by deleting a column from a symmetric $5\times 5$ matrix, and by taking $A(\phi)$ to be a symmetric $5\times 5$ matrix and $\tX$ to be $2\times 5$ (a fundamental of $O(2)$).  Then we can consider a superpotential
\be
W = \tr \Big[ \tP \Pi
\left( A(\Phi) - \tX^T \tX \right) \Big] \, ,
\ee
where
\be
\Pi = \left(\begin{array}{c|cccc}  0 &1\\ 0 &&1\\ 0 &&&1\\ 0 &&&&1 \end{array}\right) \, 
\ee
projects out the first row of $A-\tX^T\tX$.  Since $\tP$ has one fewer degree of freedom than a $5\times 5$ symmetric matrix would have, the codimension decreases by one from the corresponding $5\times 5$ case.  This general idea could be used for rectangular matrices that arise by deleting rows or columns from symmetric or antisymmetric matrices.  In these cases, the only anomaly to worry about is from the $U(1)^s$ symmetry defining the ambient variety $V$ since the nonabelian gauge group, which imposes the rank condition, is simple.

\begin{table}
\begin{tabular}{|c|c|c|c|c|c|}
\hline
$V$ & Symmetry of $A(\phi)$  &  $k$  &  $m$  &  $n$  & Group \\
\hline
\hline
$\mathbb{P}^6$ & Antisymmetric & 4 & 7 & 7  &  $USp(4)$ \\
\hline
$\mathbb{P}^7$ & None & 2 & 4 & 4 & $U(2)$  \\
\hline
$\mathbb{P}^8$ & Delete row of & 2 & 4 & 5 & $O(2)$ \\
& $5\times 5$ symmetric & & & & \\
\hline
$\mathbb{P}^9$ & Symmetric & 2 & 5 & 5 & $O(2)$\\
\hline
\end{tabular}
\centering
\caption{Assorted Calabi--Yau threefolds from \cite{Kapustka0802} and their PAXY GLSM realizations.  $V$ is the ambient toric variety, $k$ is the rank of $A(\phi)$ along the Calabi--Yau locus, and $m\times n$ is the dimension of $A(\phi)$.  $U(1)$ charges defining $V$ must (and can) be assigned in a way preserving the axial symmetry.} \label{tab:kapustka}
\end{table}

\section{Linear Determinantal Varieties}
\label{sec:linear-det-var}

We specialize to the class of linear determinantal varieties in $V=\mathbb{P}^D$ to analyze the complete phase structure of the PAX model since such an analysis in the general case seems difficult. The defining matrix is linear, $A(\phi) = \sum_{a=1}^{D+1} A^a \phi_a$, with the $A^a$ being constant $n \times n$ matrices, and $\phi_a \in \P^D$.
Since the PAX and PAXY models are dual descriptions, we will study these varieties using the more minimal PAX model, with gauge group $U(\kd) \times U(1)$ and matter content detailed in Table \ref{tab:PAX-linear}. For convenience of notation, we use $\kd = n-k$ where $k$ is the maximal rank of $A(\phi)$ along the determinantal locus.
\begin{table}
\begin{tabular}{|c|c|c|}
\hline
	& $U(\kd)$ 	& $U(1)$ \\
\hline
$X_i$	&  $\overline{\mathbf{\kd}}$ 	& $0$ \\
$P_i$ & $\mathbf{\kd}$		& $-1$ \\
$\Phi_a$	& $\mathbf{1}$		&	$1$ \\	
\hline
\end{tabular}
\centering
\caption{Chiral superfields in the GLSM and their representations under the gauge group $U(\kd)\times U(1)$. $\mathbf{\kd}$ and $\overline{\mathbf{\kd}}$ denote the fundamental and anti-fundamental representation of $U(\kd)$, and $i,j \in \{1,2,\cdots,n\}$ and $a \in \{ 1,\cdots ,D+1\}$.} \label{tab:PAX-linear}
\end{table}
The axial symmetry is anomaly-free when
\begin{equation}
D + 1 = n(n-k)\, , \label{eq:Uk-anomaly}
\end{equation}
and we will take this to be the case.  The central charge of the candidate infrared $(2,2)$ superconformal field theory will then be
\begin{equation}
\frac{c}{3} = D - (n-k)^2  =  k(n-k) - 1 \, .
\end{equation}

The scalar potential is 
\begin{eqnarray}
V_{bos} & =  & (\mbox{ D-terms })^2 + (\mbox{ F-terms })^2 + \frac{1}{2e_2^2} \tr\big([ \sigma_{U(\kd)}, \sig^\dag_{U(\kd)} ]^2\big) + 2 |\sig_{U(1)}|^2 \sum_a |\phi_a|^2\nonumber \\
&  &  +\tr \left(x \left\{ \sig_{U(\kd)},\sig^\dag_{U(\kd)} \right\} x^\dag \right) + \tr\left(p^\dag \left\{ \sig_{U(\kd)}-\sig_{U(1)},\sig^\dag_{U(\kd)}-\sigb_{U(1)} \right\} p \right)\, , \label{eq:scalar-pot}
\end{eqnarray}
with D-terms given by the classical relations
\begin{eqnarray}
U(1): &  \sum_a |\phi_a|^2 - \sum_{i,\alpha} |p_{\alpha i}|^2 = r_0 \, , \\
U(\kd): & p p^\dag - x^\dag x  = r_1 \mathds{1}_{\kd} \, ,
\label{eq:Uk-D}
\end{eqnarray}
and F-terms, which do not receive quantum corrections, given by
\begin{eqnarray}
(A^a\phi_a) x = 0\, , \qquad p (A^a\phi_a) = 0\, , \qquad \tr (pA^ax) = 0\, . \label{eq:Uk-F}
\end{eqnarray}

\subsection{Classical Moduli Space}
\label{sec:classical-moduli-space}

We first study the classical vacuum moduli space and then consider how quantum corrections modify the picture. This amounts to a study of the zero energy solutions to the scalar potential \eqref{eq:scalar-pot}. The solution space is organized into branches: the Higgs branch, where the chiral fields have expectation values that break the $U(\kd)\times U(1)$ gauge symmetry completely; Coulomb branches, where $U(1)^{\kd+1}$ gauge multiplets are massless with non-zero expectation values for corresponding $\sigma$ fields; and mixed branches where gauge bosons corresponding to a subgroup of the gauge symmetry are massless, while the commutant of the subgroup is broken.

\subsubsection{Higgs Branch}

For the sake of completeness, we repeat some of our arguments from the general discussion in Section \ref{sec:PAX}. When $r_1$ is negative, the $U(\kd)$ D-term \eqref{eq:Uk-D} implies that $x^\dag x$ is a positive-definite matrix, therefore $x$ has maximal rank $\kd$. As a consequence, the gauge group factor $U(\kd)$ is completely broken. When $r_0$ is non-zero, either the $\phi$ or the $p$ fields have non-zero expectation values depending on the sign of $r_0$. Therefore, the gauge group is broken completely for the range of parameters $r_1 < 0$, $r_0 \neq 0$, so the moduli space is a Higgs branch. By a similar argument, when $r_1>0$ the matrix $p$ has rank $\kd$, breaking the gauge group $U(\kd)\times U(1)$ to a diagonal $U(1)$, and when $r_0+\kd r_1\neq 0$ the remaining $U(1)$ is broken as well. In summary, the classical vacuum moduli space consists of a Higgs branch along 
\begin{equation}
\big\{\, (r_0, r_1) \, \big| \, r_1 < 0,\, r_0 \neq 0 \, \big\} \cup \big\{ \, (r_0, r_1) \, \big| \, r_1 > 0, \,r_0+\kd r_1 \neq 0 \, \big\}\, \subset \mathbb{R}^2 \,  . \label{eq:classical-higgs-locus}
\end{equation}
We now turn to a detailed analysis of the classical Higgs branch.

\noindent
\underline{{\bf Phase} $X_A$ ($r_0 > 0$, $r_1 < 0$):}\\
In this phase, $x$ is rank $\kd$ and $\phi$ is non-zero. The fields $(\phi , x)$ are homogeneous coordinates on $V_{\cE,\kd} = \P^D \times G(\kd,n)$ with K\"ahler parameters $(r_0,  -r_1)$. This geometric interpretation is valid when the FI parameters lie inside the classical K\"ahler cone of $ \P^D \times G(\kd,n)$: $r_0 > 0, \ r_1 < 0$.
Recall that given a Grassmannian $G(\kd,n)$, there is a natural rank-$\kd$ vector bundle known as the universal sub-bundle $\cU$ (see footnote \ref{foot:universal}), which is the analogue of the tautological line bundle $\cO_V(-1)$ over $V=\P^D$.
Then the fields $(p,\phi,x)$ form the total space of the bundle $\cO_V(-1)^{\oplus n}  \otimes  \cU$ over the base $\mathbb{P}^D \times G(\kd,n)$ with fibers $p$. Now we impose the F-term equations \eqref{eq:Uk-F}, which can be succinctly written as
\begin{equation}
E_{i\alpha} = 0\, , \qquad p_{\alpha i} \frac{\partial E_{i\alpha}}{\partial (\phi_b, \ x_{j \beta})} = 0\, , \label{eq:F-terms-jac}
\end{equation}
where $E_{i\alpha} := \phi_a A^a_{ij} x_{j\alpha} $.  The vanishing of $E_{i\alpha}$ defines a codimension $n\kd$ subvariety of $\P^D \times G(\kd,n)$
\begin{equation}
X_A = \big\{\, (\phi,x) \in \P^D \times G(\kd,n)\, \big| \ \phi_a A^a_{ij} x_{j\alpha}= 0\, \big\}\, ,
\end{equation}
which is precisely the incidence correspondence defined in Section \ref{sec:det-var}. For a generic choice of parameters $A_{ij}^a$, the variety $X_A$ is smooth, implying that the $n\kd\times (D+1+n\kd)$ (or $(D+1) \times (2D+2)$, using (\ref{eq:Uk-anomaly})) Jacobian
\begin{equation*}
\frac{\partial E_{i\alpha}}{\partial (\phi_b, \ x_{j \beta})}\, 
\end{equation*}
has rank $n\kd$, equal to the codimension of $X_A$ as a subvariety of $\P^D \times G(\kd,n)$. The F-terms \eqref{eq:F-terms-jac} thus force $p =0$, yielding a vacuum moduli space given by the smooth variety $X_A$.

\noindent
\underline{{\bf Phase} $X_{A^T}$ ($r_0+\kd r_1 > 0$, $r_1 > 0$):}\\
We consider a linear combination of the D-terms 
\begin{equation}
\sum_a |\phi_a|^2 - \sum_{i,\alpha} |x_{i \alpha }|^2 = r_0 + \kd r_1\, .
\end{equation}
In this phase, $p$ has rank $\kd$ and $\phi$ is non-zero. We redefine the $\C^*\times GL(\kd,\C)$ action as
\begin{equation}
(\phi,p) \rightarrow (\lambda \, \phi, M p)\, , \qquad M\in GL(k,\C), \ \lambda \in \C^*\, .
\end{equation}
The fields $(x,\phi,p)$ form the total space of the vector bundle $\cO_V(-1)^{\oplus n} \otimes \cU$, with $(\phi,p) \in \P^D\times G(\kd,n)$ forming the base with K\"ahler class  $(r_0+\kd r_1,r_1)$, and $x$ the coordinates of the fiber. An analysis similar to the previous one shows that $x=0$ and that the vacuum moduli space is the smooth incidence correspondence
\begin{equation}
X_{A^T} = \big\{\ (\phi, p) \in \P^D \times G(\kd,n)\ \big|\  \phi_a A_{ij}^a p_{\alpha i}= 0\ \big\}\, .
\end{equation}
As the name suggests, $X_{A^T}$ denotes the incidence correspondence constructed using the transposed matrix $A(\phi)^T$.

\noindent
\underline{{\bf Phase} $Y_A$ ($r_0 < 0$, $r_1 < 0$):}\\
Here, $x$ must have rank $\kd$ and $p$ is non-zero with $1\leq \rank(p) \leq \kd$. These fields are homogeneous coordinates on the total space of the projectivized vector bundle
\begin{equation}
\P(\cU^{\oplus n})  \stackrel{\pi}{\longrightarrow} G(\kd,n) \, .
\label{eq:bundle-gr-phase}
\end{equation}
The K\"ahler class of the base is  $-(r_0+\kd r_1)$, while the class of the fiber is $-r_0$. The F-terms \eqref{eq:Uk-F} can be written as
\begin{equation}
E_a = 0\, , \qquad \phi_a\, \frac{\partial E_a}{\partial (x_{i\alpha}, p_{\beta j})} = 0\, ,
\end{equation}
where $E_a := \tr(p A^a x)$. Again, for a generic choice of parameters $A_{ij}^a$ the F-terms impose $\phi=0$ and the moduli space is the variety
\begin{equation}
Y_A := \big\{\ (x,p) \in \P(\cU^{\oplus n}) \ \big| \ \tr(pA^ax) = 0 \ \big\} \, .
\end{equation}
$Y_A$ can be viewed as an incidence correspondence as follows: the matrix $\mathcal{Y}^a_{i\alpha} := A^a_{ij}x_{j\alpha}$ is a $(D+1)\times (D+1)$ matrix and along $Y_A$\,, $p$ is in the kernel of $\mathcal{Y}$.  Thus, we have a determinantal variety in $G(\kd,n)$ defined by the hypersurface $\det(\mathcal{Y}) = 0$ or, equivalently, as the locus $\rank(\mathcal{Y}) \leq D$ with singularities resolved by the incidence correspondence $Y_A$.  Over a generic point of the hypersurface, the fiber is just a point (after accounting for the $\C^*$ action), but when $\rank(\mathcal{Y})$ drops further to rank $(D-l)$ for $0<l\leq D$, then the kernel becomes a $\P^l$ and corresponds to a resolution of the singularity.

\noindent
\underline{{\bf Phase} $Y_{A^T}$ ($r_0+\kd r_1 < 0$, $r_1 > 0$):}\\
In this phase, $p$ has rank $\kd$ and $1 \leq \rank(x) \leq \kd$. If we redefine the $\C^*\times GL(\kd,\C)$ action as
\begin{equation}
(x,p) \rightarrow (M^{-1} \lambda \, x, M  p) \, , \qquad M\in GL(\kd,\C), \ \lambda \in \C^*\, ,
\end{equation}
the analysis is identical to the previous case with the roles of $p$ and $x$ switched. The ambient space is again the projective bundle $\P(\cU^{\oplus n})$ over $G(\kd,n)$, where the K\"ahler class of the base and fiber are  $(-r_0, -(r_0+\kd r_1))$. The moduli space is a smooth incidence correspondence
\begin{equation}
Y_{A^T} = \big\{\ (p,x) \in \P(\cU^{\oplus n})\ \big|\ \tr(pA^ax)=0 \ \big\}\, .
\end{equation}
Note that when it is impossible for $l$ to take a nonzero value, $Y_A$ and $Y_{A^T}$ will be isomorphic.  We will elaborate on this possibility when we discuss the potential phase boundary along $r_1=0$, $r_0<0$.

 \begin{figure}[t]
 \centering
 \includegraphics[width=1.5in]{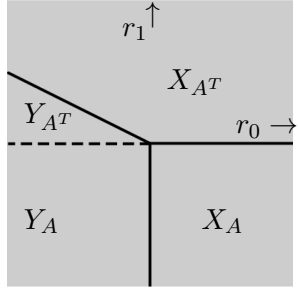}
 \put(-48,75){$X_{A^T}$}
 \put(-102,62){$Y_{A^T}$}
 \put(-102,22){$Y_A$}
 \put(-35,22){$X_A$}
 \put(-22,58){$r_0 \rightarrow$}
 \put(-65,95){$r_1$}
 \put(-56,100){$\uparrow$}
\caption{The classical vacuum moduli space of the GLSM as a function of the FI parameters $(r_0,r_1)$: the Higgs branch is shaded in grey, and the Coulomb branch locus is the thick solid lines.  The phase boundary between $Y_{A^T}$ and $Y_A$ only exists when $\lfloor\sqrt{nk-1}\rfloor \geq k+1$ and $\kd \geq 2$.  In particular, for $(k,n)=(2,4)$ or $(4,5)$ it is absent.  The phase boundary between $Y_{A^T}$ and $X_{A^T}$ is defined by $r_0 + \kd r_1 = 0$.}
 \label{fig:phases-classical}
 \end{figure}

To summarize, the classical moduli space when equation \eqref{eq:classical-higgs-locus} is satisfied is a Higgs branch and has the geometry of an incidence correspondence, depicted in Figure \ref{fig:phases-classical}.  Each of the varieties $X_A$, $X_{A^T}$, $Y_A$, and $Y_{A^T}$, is smooth for a generic choice of the coefficients $A_{ij}^a$.  Separating these phases are loci which are not pure Higgs branches: one or more gauge multiplets are massless with the corresponding $\sigma$ fields obtaining non-zero expectation values.  As these boundaries are approached, the geometric description of the moduli space as a Calabi--Yau breaks down; for example, certain holomorphic cycles in the Calabi--Yau may shrink to zero volume.  In these regimes, we will have to account for quantum effects and we do so in Section \ref{sec:quantum-phase}.

\subsubsection{Mixed Branches}
\label{sec:classical-mixed-branches}

We describe below mixed branches where gauge bosons in a subgroup of $U(\kd)\times U(1)$ are massless while the commutant is broken by expectation values of the chiral multiplets.

\noindent
\underline{\bf $r_1 < 0$, $r_0=0$:}\\
$x$ has rank $\kd$, completely breaking $U(\kd)$.  Since $r_0=0$, the D-terms force both $\phi$ and $p$ to be zero, or both to be non-zero; the F-terms \eqref{eq:Uk-F} then force $\phi = 0$ and $p = 0$.  This leaves a massless $U(1)$ Coulomb branch with $\sig_{U(1)} \neq 0,$ $\sig_{U(\kd)}=0$.

\noindent
\underline{$r_1 > 0$, $r_0+\kd r_1=0$:}\\
$p$ has rank $\kd$, breaking $U(\kd)\times U(1)$ down to a diagonal $U(1)$ gauge group, leaving a massless $U(1)$ Coulomb branch parameterized by $\sig_{U(\kd)} = \sig_{U(1)} \mathbf{1}_{\kd \times \kd}$, $\sig_{U(1)} \neq 0$.

\noindent
\underline{\bf $r_1 = 0$, $r_0> 0$}:
\\
The D-terms \eqref{eq:Uk-D} impose $\phi\neq 0$ and $pp^\dagger = x^\dagger x$, which allow for a mixed Coulomb--Higgs branch with 
\begin{equation*}
\sig_{U(1)} = 0\, , \quad \sig_{U(\kd)} = \mathrm{diag}\, \{\sig_1, \ldots, \sig_l, 0, \ldots, 0\, \}\, , \quad \rank(x) = \rank(p) \leq \kd-l\, , \quad 0 < l \leq \kd\, .
\end{equation*}
We now analyze the effect of the F-terms on this branch of solutions. The F-terms are satisfied when $l=\kd$ since we have $x=p=0$. This results in a mixed Coulomb--Higgs branch with $\phi \neq 0$ and massless $U(1)^\kd$ gauge multiplets.

When $0 < l < \kd$, the F-terms cannot be satisfied generically.  To see this, we first define an $n\times n$ matrix of ``meson'' fields $M := xp$, in terms of which the F-terms become
\begin{equation}
(A^a\phi_a) M = 0\, , \qquad M (A^a\phi_a) = 0\, , \qquad \tr(A^a M) = 0\, . \label{eq:Ft-meson}
\end{equation}
For generic parameters $A^a$, $\phi \neq 0$ implies that $\rank(A^a \phi_a) \geq n - \fl{\sqrt{n\kd-1}}$, where $\lfloor \cdots \rfloor$ denotes greatest integer less-than-or-equal-to its argument.\footnote{The rank $m$ locus is codimension $(n-m)^2$. The minimal rank is determined by demanding that the codimension  is less than the dimension of the ambient space, $D = n\kd - 1$.} 
Equivalently, the dimension of the null space of $A^a\phi_a$ is at most $\fl{\sqrt{n\kd-1}}$, so the first two conditions in \eqref{eq:Ft-meson} constrain the number of degrees of freedom in $M$ to be at most $(\fl{\sqrt{n\kd-1}})^2$. The condition $\tr(A^a M) = 0$ imposes $n\kd-1$ additional independent conditions on $M$, and since these conditions are homogeneous, the only solution generically is $M=0$. The D-term $pp^\dag = x^\dag x$ then requires that $x=0$ and $p=0$.  Since this is incompatible with the assumption that $0 < l < \kd$, this mixed Coulomb--Higgs branch is lifted by the F-terms.

\noindent
\underline{\bf $r_1 = 0$, $r_0< 0$}:
\\
The D-terms \eqref{eq:Uk-D} again impose $pp^\dagger = x^\dagger x$, and $r_0<0$ implies $p\neq 0$.  Note that they imply that $xp \neq 0$.  This allows for a potential mixed Coulomb--Higgs branch along the locus
\begin{equation}
\sig_{U(1)} = 0\, , \qquad \sig_{U(\kd)} = \mathrm{diag}\, \{\sig_1, \ldots, \sig_l, 0, \ldots, 0\, \}\, , \qquad \rank(x) = \rank(p) \leq \kd-l\, , 
\end{equation}
with $1 \leq l < \kd$ (note that this branch exists only when $\kd \geq 2$). We can solve the F-term condition $\tr(A^a xp) = 0$ through the ansatz $xp = B^\ac \phic_\ac$, $\ac = 1, \ldots, nk$, where the matrices $B^\ac$ are generic and satisfy $\tr(B^\ac A^a)=0$ (we will address when $xp=B^\ac \phic_\ac$ can be nonzero in the next paragraph).
In terms of the new variables $\phic_\ac$, the remaining F-term equations can be recast as
\begin{equation}
(B^\ac \phic_\ac) N = 0\, , \qquad N (B^\ac\phic_\ac) = 0\, ,
\end{equation}
where we defined the matrix $N := A^a\phi_a$.  Since $xp \neq 0$ implies $\phic\neq 0$, this ``dual'' formulation of the F-term constraints allows us to conclude that we must set $N=0$, appealing to our arguments from the analysis of the previous branch.  

Returning to the ability to set $xp = B^\ac \phic_\ac$, since $1 \leq \rank(B^\ac \phic_\ac) = \rank(xp) \leq \kd-l$, we see that the moduli space will be $Z(B,\kd-l)\subset\mathbb{P}^{n k-1}$.  Requiring $Z(B,\kd-l)$ to be nonempty implies that its codimension $(k+l)^2$ does not exceed $(nk-1)$, so  $k+1 \leq k+l \leq \lfloor \sqrt{n k-1}\rfloor$.  Therefore, such a mixed branch is possible whenever
\begin{equation}
k+1 \leq \fl{\sqrt{nk-1}}\ \mbox{  and  }  \  \kd \geq 2\, . \label{eq:mixed-branch-condition}
\end{equation}

\subsubsection*{Pure Coulomb Branch}
\noindent
\underline{$r_1 = 0$, $r_0=0$:}\\
At the origin in the parameter space, the D-terms allow for two possibilities: either all the chiral fields $\phi,~x,$ and $p$, are simultaneously zero, or simultaneously non-zero.  The F-terms force $\phi=0,~x=0,$ and $p=0$. The commutator $\tr([\sig,\sig^\dag]^2)$ breaks the gauge group down to the maximal abelian subgroup $U(1)^{\kd+1}$ at a general point in the $\sigma$ field space. Along special subloci in $\sigma$-space, where eigenvalues become degenerate, the gauge group can be enhanced to a nonabelian factor.

\subsection{Quantum Moduli Space of the GLSM}
\label{sec:quantum-phase}

In this section, we will study the quantum corrections to the classical moduli space. The $(2,2)$ GLSM flows to an interacting $(2,2)$ superconformal field theory in the infrared, and the complexified FI parameters $(t_0,t_1)$ in the gauge theory correspond to exactly marginal deformations of the infrared superconformal field theory.  For asymptotically large values of the FI parameters, far from the classical Coulomb branch loci, the theory is weakly coupled so we expect the classical description to be a good approximation: the GLSM flows to a nonlinear sigma model with target space an incidence correspondence. As the FI parameters are tuned closer to the Coulomb (or mixed) phases, quantum effects become important and we gradually lose the geometric interpretation. Near the classical Coulomb branches where a non-compact direction for $\sigma$ appears to emerge, theta angles can be turned on to lift these noncompact directions via a constant background electric field, allowing one to smoothly interpolate from one Higgs phase to another --- in other words, the singular Coulomb branches, which appear to be real codimension one in Figure \ref{fig:phases-classical}, are actually \emph{complex} codimension one in the complexified FI parameter space \cite{Witten:1993yc}.

The classical analysis does not tell us precisely where the singular loci are in the quantum-corrected complexified K\"ahler moduli space of the Calabi--Yau, however the GLSM actually allows us to compute the singular loci.  This can then be compared with predictions from mirror symmetry when available, though mirror symmetry for generic determinantal Calabi--Yau varieties is still an open problem and examples are scarce.  Our computations in this section will thus serve as a test of any future mirror proposals.  In order to compute the quantum corrected singular locus, where the Coulomb branch emerges, we adopt the methods developed in \cite{Witten:1993yc, Hori:2006dk}. The Coulomb branch is associated with non-zero expectation values for the $\sigma$ fields, so we will study the effective field theory obtained by giving large, slowly-varying expectation values to them and then integrating out all the massive fields to obtain an effective theory for the \textsc{vev}s of $\sig$.  For \emph{generic} values of the $\sigma$ \textsc{vev}s, the effective field theory consists of $(\kd+1)$ twisted chiral fields, corresponding to the $U(1)^{\kd+1}$ Coulomb branch, with an effective twisted superpotential.  We will see that the twisted superpotential only has extrema when $(t_0,t_1)$ lie along a complex codimension one divisor, giving us a component of the singular locus where a noncompact direction in $\sigma$ space develops.  Each Coulomb or mixed branch must be individually studied to determine its contribution to the singular locus, if any.  We provide evidence that the branches with nonabelian gauge symmetry are absent from the quantum theory.  Therefore we argue that the singular locus in FI parameter space --- which we explicitly compute --- corresponds, in general, to a pure Coulomb branch or a mixed Coulomb--Higgs branch.

\subsubsection{Pure Coulomb Branch: $U(\kd)\times U(1) \rightarrow U(1)^{\kd+1}$}

We study the effective field theory in a classical background where all the $\sigma$ fields have large expectation values. In the classical bosonic potential \eqref{eq:scalar-pot}, the commutator-squared term $\tr\big([\sigma,\sigma^\dag]^2\big)$ vanishes only when $\sig$ and $\sig^\dag$ commute, which implies that they are diagonalizable. We take the ansatz
\begin{equation}\label{eq:gen-sig-vevs}
  \sig_{U(\kd)} = \mathrm{diag} \!\left\{ \sig_1, \ldots, \sig_{\kd} \right\} \ ,  \qquad \sig_{U(1)} = \sig_0\ . 
\end{equation}
We will assume that the expectation values of the  $\sig_\alpha$ along the Cartan directions are large, slowly varying, and (for the moment) completely generic:
\begin{equation}
\sig_\alpha \neq \sig_\beta\, , \qquad \sig_0 \neq 0\, , \qquad \sig_\alpha \neq 0\, ,\qquad \sig_0 \neq  \sig_\alpha\, , \qquad \alpha,\beta = 1,\ldots,\kd\, . \label{eq:sig-generic}
\end{equation}
The gauge group is then broken down to the maximal abelian subgroup $U(1)^{\kd+1}$,
and the $W$-bosons have large masses of order $|\sig_\alpha-\sig_\beta|$.  The chiral multiplets also get large masses as shown in Table \ref{tab:Coulomb-charges}.

\begin{table}
\centering
\begin{tabular}{|c|c|c|c|c|c|}
\hline
		 &$U(1)_0$ 	& $U(1)_{\beta=1,\ldots,\kd}$  &  mass${}^2$  \\
\hline
$X_{\alpha,i}$	 & $0$ 		&  $-\delta_{\alpha\beta}$	&	 $2\big| \sig_\alpha\big|^2$ \\
$P_{i,\alpha}$	 & $-1$ 		& $\delta_{\alpha\beta}$	&	$2\big| \sig_0-\sig_\alpha\big|^2$ \\
$\Phi_a$	 & $1$ 	& $0$	& $2\big| \sig_0 \big|^2$ \\
\hline
\end{tabular}
\caption{Chiral superfields, masses, and charges under $U(1)^{\kd+1}$ on the Coulomb branch.}
\label{tab:Coulomb-charges}
\end{table}

There is a simple formula for the effective twisted superpotential obtained by integrating out the heavy gauge bosons and chiral multiplets \cite{Witten:1993yc, summing}
\begin{equation}
2\sqrt{2}\ \tW_{\mathit{eff}}(\Sig)  =   \Sig_0 \left( it_0 + \frac{n}{2\pi} \log \prod_{\alpha=1}^\kd  \frac{\Sig_\alpha-\Sig_0}{\Sig_0} \right)  + \sum_{\alpha=1}^\kd \Sig_\alpha \left( it_1 - \frac{n}{2\pi} \log \left[ \frac{\Sig_0-\Sig_\alpha}{\Sig_\alpha}\right]\right) \label{eq:CB-eff-W} \, .
\end{equation}
The equations that follow from this effective superpotential are
\begin{equation} \label{eq:CB-eq-1}
q_0^{-1}\,=\,\left(\prod_{\alpha=1}^{\kd} \frac{(\Sig_\alpha-\Sig_0)}{\Sig_0}\right)^n \ , \qquad
q_1\,=\,\left(\frac{\Sig_0-\Sig_\alpha}{\Sig_\alpha}\right)^n \ , \qquad \alpha = 1,\ldots,\kd \  , 
\end{equation}
where $q_0 = \exp(2\pi i t_0)$ and $q_1 = \exp(2\pi i t_1)$ are coordinates respecting the periodicity of the theta angles.  Note that the equations are invariant under the scaling $\Sig_0 \rightarrow \lambda \Sig_0, \ \Sig_\alpha \rightarrow \lambda \Sig_\alpha$, so there will be a non-compact Coulomb branch direction whenever a solution to \eqref{eq:CB-eq-1} exists. In terms of the combinations
\begin{equation}
\textsc{x}_\alpha := \frac{\Sig_0-\Sig_\alpha}{\Sig_\alpha}\, , \qquad \alpha=1 , \ldots, \kd \, ,
\end{equation}
equation \eqref{eq:CB-eq-1} then simplifies to
\begin{equation}
  \textsc{x}_\alpha^n = q_1\ , \quad \alpha = 1,\ldots,\kd\  , \qquad
  \prod_{\alpha=1}^\kd (1+\textsc{x}_\alpha)^{n} = (-1)^{n\kd}\, q_0\, q_1^\kd\ . \label{eq:CB-eq-2}
\end{equation}
The genericity conditions \eqref{eq:sig-generic} require that
\begin{equation}
\textsc{x}_\alpha \neq 0 , \, -1 , \qquad \mbox{and} \qquad \textsc{x}_\alpha \neq \textsc{x}_\beta  \quad \mbox{for} \quad \alpha \neq \beta \, .
\end{equation}
We have a system of $(\kd+1)$ independent equations in \eqref{eq:CB-eq-2} for the $\kd$ variables $\textsc{x}_\alpha$, so generically they only have a solution along a codimension one locus in FI parameter space.  Introducing a variable $\xi$ such that $\xi^n=q_1$, we can write
\begin{equation}
\textsc{x}_\alpha = \omega^{n_\alpha} \xi  \quad \mbox{with }  \ \omega := e^{\frac{2\pi i}{n}}\, , \qquad n_\alpha \neq n_\beta \mbox{ for } \alpha\neq \beta\, .
\end{equation}
The singular locus therefore consists of a union of irreducible components $\cC_{n_\alpha}$ defined by the parameterization
\begin{equation}
\big(q_0(\xi),q_1(\xi)\big) = \left( (-1)^{n\kd} \prod_{\alpha=1}^\kd \frac{(1 + \omega^{n_\alpha} \xi)^n}{\xi^n}\ ,\ \ \xi^n\right)\, . \label{eq:sing-locus-param}
\end{equation}
Since the parameterization is in terms of rational functions, each component $\cC_{n_\alpha}$ is a genus zero rational curve in the $(q_0,q_1)$ plane. A definition of the curve $\cC_{n_\alpha}$ in the implicit form $f_{n_\alpha} (q_0,q_1) = 0$ can be obtained using the \emph{resultant} as \cite{Rational-Curves-book}
\begin{equation}
f_{n_\alpha} (q_0,q_1) = \mathrm{res}_\xi \left( \xi^{n\kd} q_0 - (-1)^{n\kd}  \prod_{\alpha=1}^\kd (1 + \omega^{n_\alpha} \xi)^n \, , \  q_1 - \xi^n\right)\, .  \label{eq:sing-locus-implicit}
\end{equation}
A general formula for the defining equation seems difficult to obtain, but for specific values of $\kd$ and $n$ the singular locus is easily determined and we compute it for the examples in Section \ref{sec:examples}.

The number of components of the singular locus is the number of $\{n_\alpha\}$ up to automorphisms.  The assumption of genericity \eqref{eq:sig-generic} implies that $n_\alpha \neq n_\beta$ if $\alpha\neq\beta$.  Furthermore, the ansatz \eqref{eq:sig-generic} leaves $S_\kd \subset U(\kd)$ unbroken, which we can fix by choosing $0\leq n_1 < n_2 < \ldots < n_\kd < n$.  This would lead to ${ n \choose \kd}$ loci, but there is an additional invariance that relates $C_{n_\alpha}$ to $C_{n'_\alpha}$ if $n'_\alpha=n_\alpha + \delta$ for integral $\delta$ (they are related by a change of parameterization $\xi \rightarrow \omega^\delta \xi$).  This is not such a straightforward problem to solve because shifts of certain configurations of the $\{n_\alpha\}$ can be the same as elements of $S_\kd$ (otherwise, one could simply fix $n_1 = 0$ and obtain ${n-1 \choose \kd-1}$, which would be a large overcounting).  

One can rephrase the problem into counting the number of degree $(n-\kd)$ polynomials in $\kd$ variables (again, ${n-1 \choose \kd-1}$ in total) that are invariant under cyclic permutations $\mathbb{Z}_{\kd}$.  In particular, we can choose coordinates to diagonalize the cyclic $\mathbb{Z}_\kd$ so that it acts via multiplication by the $\kd$th roots of unity.  Embedding this $\mathbb{Z}_{\kd}$ into a $U(1)$ so that the charge of the $i$th coordinate is $i$, where $i=0,\ldots,\kd-1$, we are interested in counting degree $(n-\kd)$ polynomials whose $U(1)$ charges are integer multiples of $\kd$.  This becomes a classic problem in the theory of algebraic invariants, and one can find a representation for this answer in \cite{hilbert1993theory}.  There,
\be
\omega_{\kd-1}(n-\kd,p) := \oint \frac{dx}{2\pi i x^{p+1}}\ \frac{(1-x^{\kd}) (1-x^{\kd+1}) \cdots (1-x^{n-1})}{(1-x)(1-x^2)\cdots (1-x^{n-\kd})} 
\ee
counts the number of charge $p$ degree $(n-\kd)=k$ polynomials in $\kd$ variables.  In our case, we want to sum over $p = N\kd$ for integer $N$, where $N$ runs from $0$ to $\lfloor k(1-\kd^{-1}) \rfloor$.  For example, when $\kd=2$ and $n=4$ we have
\be
\omega_1(2,0) + \omega_1(2,2) = 1 + 1 = 2 \, ,
\ee
as confirmed by the explicit example to follow.  

So far, we have considered directions in field space that satisfy \eqref{eq:sig-generic}. When some of the eigenvalues $\sig_\alpha$ coincide, the low-energy effective theory has a nonabelian gauge symmetry. These special directions could, in general, yield additional noncompact directions that would correspond to additional components of the singular locus. We will examine this possibility in Section \ref{sec:CB-nonabelian}.

\subsubsection{Mixed Coulomb--Higgs branch}
\label{sec:CB-mixed}

\noindent
\underline{$r_0 > 0$, $r_1 = 0$:}\\
Classically, $\phi$ must be nonzero while $x=0$ and $p=0$.  
We consider the GLSM in a background
\begin{equation} 
\sig_{U(\kd)} =
  \mathrm{diag}\! \left\{\sig_1, \ldots, \sig_{\kd}\right\} \ ,  \qquad  \sig_\alpha \neq 0\ ,  \qquad 
  \sig_\alpha  \neq \sig_\beta \textrm{~~for~~} \alpha\neq\beta \  ,
\end{equation}
which breaks the $U(\kd)$ factor to the maximal abelian subgroup $U(1)^\kd$. In this background, we can integrate out the the massive $W$-bosons and the massive chiral fields $x$ and $p$. We now look for a Higgs branch in the $U(1)_0$ gauge theory with charged chiral fields $\Phi_a$, and couplings that depend on the background fields $\sig_\alpha$. The effective twisted superpotential is given by
 \begin{equation}
 \label{eqn:mixed-eff-tW}
 2\sqrt{2}\ \tW_{\mathit{eff}}  =   \Sigh_0 \left( it_0 + \frac{n}{2\pi} \log  \frac{ \prod_{\alpha=1}^\kd(-\Sigh_0+ \Sig_\alpha)}{(2^{-1/2}\mu)^\kd} \right)  + \sum_{\alpha=1}^\kd \Sig_\alpha \left( it_1 - \frac{n}{2\pi} \log  \frac{\Sigh_0-\Sig_\alpha}{\Sig_\alpha} \right) \, ,
 \end{equation}
which generates an effective $U(1)_0$ D-term given by
\be
\label{eqn:mixed-eff-D0}
\sum_a |\phi_a|^2 =  r^{\mathit{eff}}_0(\sigma,\mu) := r_0 - \frac{n \kd}{2\pi} - \frac{n}{4\pi} \sum_{\alpha=1}^\kd \log \bigg( \frac{2\big|\sig_\alpha\big|^2}{|\mu|^2} \bigg)\, , \qquad  r_1^{\mathit{eff}}(\sig,\mu) = r_1 = 0 \, ,
\ee
along a potential Higgs branch where $\sigh_0=0$.  Here $\mu$ represents some energy subtraction scale.  In \eqref{eqn:mixed-eff-tW}, we have used the hatted symbol $\Sigh_0$ to emphasize that it is not a background field.  Note that $r_1$ does not run since we have integrated out fields pairwise with equal and opposite charges. As we flow to the infrared, $\mu\rightarrow 0$ and $r_0^{\mathit{eff}} \rightarrow -\infty$, forcing $\phi = 0$.  Therefore, the Higgs branch we are looking for has been lifted by quantum effects. The fields $\phi_a$ are massive and we must integrate them out as well, returning us to the result we found when integrating out $\phi$, $x$, and $p$, in the previous section.  We thus have no additional singular loci from this mixed Coulomb--Higgs branch.

\noindent
\underline{$r_0 < 0$,  $r_1 = 0$:}\\
Classically, we have a mixed Coulomb--Higgs branch with 
\begin{equation}
\sig_{U(1)} = 0\, , \qquad  \sig_{U(\kd)} = \mathrm{diag}\! \left\{\sig_1, \ldots, \sig_l, 0, \ldots, 0 \right\} \, , \qquad   \sig_\alpha \neq 0 \, , \quad \sig_\alpha\neq \sig_\beta \textrm{~~for~~} \alpha\neq\beta \, ,
\end{equation}
and with $\phi=0$, $\rank(x)=\rank(p) \leq \kd-l$, where $1 \leq l \leq \min\{\kd-1, \fl{\sqrt{nk-1}}-k\}$. Such a branch of solutions exists only when \eqref{eq:mixed-branch-condition} is satisfied. 

We consider the dynamics of the GLSM in the background field
\begin{equation}
\sig_{U(\kd)} = \mathrm{diag} \!\left\{ \sig_1, \ldots, \sig_l, 0,\ldots,0 \right\} \ , \qquad \sig_\alpha \neq 0 \ ,  \qquad \sig_\alpha\neq \sig_\beta \textrm{~~for~~} \alpha\neq\beta \ .
\end{equation}
In this background field a subset of the $x$ and $p$ fields are massive, in addition to the $W$-bosons, and we integrate these out. We denote the remaining fields using hatted symbols for the sake of clarity. Let $\hx$ and $\hp$ be $n\times (\kd-l)$ and $(\kd-l)\times n$ sub-matrices of $x$ and $p$, respectively. Along with the $\phih_a$, these fields are all charged under the unbroken $U(1)_0\times U(\kd-l)$ gauge symmetry.
\bea
2\sqrt{2}\ \tW_{\mathit{eff}}  &=&   \Sigh_0 \left( it_0 + \frac{n}{2\pi} \log  \frac{ \prod_{\alpha=1}^l  (\Sig_\alpha-\Sigh_0)}{(2^{-1/2}\mu)^l} \right)  + \sum_{\alpha=1}^l \Sig_\alpha \left( it_1 - \frac{n}{2\pi} \log  \left[\frac{\Sigh_0-\Sig_\alpha}{\Sig_\alpha}\right] \right)   \nonumber \\
&& + it_1 \tr \Sigh_{U(\kd-l)} \, . ~
\eea
On a potential Higgs branch, $r^{\mathit{eff}}_0$ will flow to more negative values, while $r_1$ is not renormalized. As a consequence the classical analysis of the mixed branch, carried out in section \ref{sec:classical-mixed-branches}, is a good approximation at low energies. We have a Higgs branch with $\phih=0$, while $\hx$ and $\hp$ satisfy $\tr(\hp A^a \hx) = 0$.

Now, we solve for the dynamics of the remaining slowly-varying background fields $\Sig_\alpha$. On the Higgs branch of the $U(1)\times U(\kd-l)$ theory ($\Sigh_0=0$, $\Sigh_{U(\kd-l)}=0$), extremizing the effective twisted superpotential with respect to $\Sig_\alpha$ implies
\begin{equation}
 \left( it_1 - \frac{n}{2\pi} \log  (-1) \right) = 0 \, ,
\end{equation}
thereby resulting in a noncompact $U(1)^l$  mixed Coulomb--Higgs branch whenever
\be
\label{eq:mixed-divisor}
q_1 = (-1)^n \, . 
\ee
We reiterate that this singular divisor only exists when
\begin{equation}
k+1 \leq \fl{\sqrt{nk-1}}\ \mbox{  and  }  \  \kd \geq 2\,  
\end{equation}
are satisfied.

\noindent
\underline{$r_0 = 0$, $r_1 < 0$:}\\
Classically $\sig_{U(1)} \neq 0$, while $\rank(x)=\kd$, which breaks the $U(\kd)$ factor completely.  Quantum mechanically, in a background with $\sig_{U(1)} \neq 0$, we integrate out the heavy fields $p$ and $\phi$ to obtain the following effective D-term equation along a potential Higgs branch
\begin{equation}
\tr(x^\dag x) = -r^{\mathit{eff}}_1(\sig,\mu) := -r_1 - \frac{n \kd}{4\pi} \log \left| \frac{\sig_{U(1)}}{\mu} \right|^2\, .
\end{equation}
Thus, $r_1^{\mathit{eff}}$ flows to $-\infty$ in the infrared, forcing $x$ to have vanishing \textsc{vev} and become massive.  This branch is completely lifted by quantum effects.

\noindent
\underline{$r_0 + \kd r_1 = 0$,  $r_1 > 0$:}\\
Classically, $p$ has rank $\kd$ and breaks the $U(\kd)\times U(1)$ down to a diagonal $U(1)$.  The analysis is essentially identical to that of the previous case with the roles of $p$ and $x$ interchanged.  The conclusion is the same: no new singular loci appear.

\noindent
\subsubsection{Additional Singularities from Nonabelian Dynamics}
\label{sec:CB-nonabelian}

So far, we have studied the phase structure of the FI parameters where the gauge theory is in a Higgs phase, and we have studied the part of the phase structure where other branches become possible, but we have only allowed for large distinct \textsc{vev}s of (\ref{eq:sig-generic}), giving all the chiral superfields and $W$-bosons large masses.  To complete the search for other possible noncompact directions, we must fill in the gaps at these latter loci and study what happens when $\sigma$ have small \textsc{vev}s.

We will closely follow the argument constructed by Hori and Tong \cite{Hori:2006dk}.  First, note that a noncompact direction signals a singularity in the quantum K\"ahler moduli space of the Calabi--Yau; since this is independent of the complex structure and, therefore, the superpotential, we can set $W=0$ to obtain a simpler GLSM  (denoted GLSM2) whose quantum K\"ahler moduli space contains the singularities of the original GLSM --- i.e., a singularity in FI parameter space of GLSM must be a singularity in FI parameter space of GLSM2, but it can happen that GLSM2 contains additional singularities (geometrically, the ambient space may contain singularities that the subvariety defined by $W$ does not intersect).  In GLSM2, we can compute the Witten index on the Higgs branch by turning on twisted masses for the chiral fields, regulating flat directions and allowing the Witten index to be computed as the Euler characteristic of the vacuum manifold.\footnote{Twisted masses can be turned on for a chiral superfield only when phase rotations of that field are a global symmetry.  This is why we cannot turn on generic twisted masses in the original GLSM and why we drop the superpotential in defining GLSM2.}  The twisted masses also regulate the flat directions on the Coulomb branch and turn the vacua into a set of points, enabling us to compute the Witten index.  Finding agreement between Higgs and Coulomb branches supports the hypothesis that there are no vacua that we overlooked in the strongly coupled regions and, therefore, no reason to anticipate additional singular loci --- of course, we cannot rule out the possibility that vacua appear and cancel in bosonic/fermionic pairs.

We set the superpotential to zero and turn on twisted masses $\tm_p$ for $p$.  For large values of the FI parameters, $|r_{0,1}| \rightarrow \infty$, the moduli space is in the Higgs branch. For $r_0 > 0$ and $r_1 < 0$, and below the energy scale $\tm_p$, GLSM2 flows to a nonlinear sigma model with target space $G(\kd,n)\times \P^D$. The Witten index of the nonlinear sigma model is the Euler characteristic of the target space \cite{Witten:1982df}, which in our case is
\begin{equation}
\chi\big(\P^{D} \times G(\kd,n)\big) = \chi(\P^{D}) \, \chi(G(\kd,n)) = (D+1) \bin{n}{\kd} \, . \label{eq:vac-nlsm}
\end{equation}
Moreover, the target space has no odd cohomology, so this is the total number of (bosonic) vacua.

Now consider the theory with FI parameters $r_0$ and $r_1$ non-zero but small.  In this regime, the Higgs description is not reliable but the Witten index remains the same. Classically, when twisted masses are present the $\sigma$ fields have non-zero expectation values and the chiral fields are massive. For large twisted masses, the classical picture is approximately valid so we expect to find the vacua in the field space spanned by the $\sigma$. The effective twisted superpotential for the light $\Sig_{0}$ and $\Sig_\alpha$ for large twisted masses is
\bea
2\sqrt{2} \ \tW_{\mathit{eff}}(\Sig)  &=&  it_0\Sig_0 + it_1\sum_{\alpha=1}^\kd \Sig_\alpha
+\frac{n}{2\pi}\sum_{\alpha=1}^\kd \Sig_\alpha\log \left(- \frac{\sqrt{2}\Sig_\alpha}{\mu}\right)    -\frac{n\kd }{2\pi} \Sig_0 \log \left( \frac{\sqrt{2} \Sig_0}{\mu} \right)        \nonumber \\
&&  -\frac{n}{2\pi}\sum_{\alpha=1}^\kd \left( -\Sig_0 + \Sig_\alpha-\tm_p\right)\log \left( \frac{\sqrt{2}(-\Sig_0 + \Sig_\alpha-\tm_p)}{\mu} \right)  \, ,
\eea
and the equations for the vacua are
\begin{equation} \label{eq:CB-twisted-eqns}
q_0^{-1}\,=\,\prod_{\alpha=1}^\kd \left( \frac{-\Sig_0+\Sig_\alpha-\tm_p}{\Sig_0}\right)^n\  ,  \qquad
q_1\,=\,\left( \frac{-\Sig_0+\Sig_\alpha-\tm_p}{-\Sig_\alpha}\right)^n \ , \quad \alpha = 1,\ldots,\kd \ . 
\end{equation}

The equations have a scale invariance $\Sig_{0,\alpha} \rightarrow \lambda \Sig_{0,\alpha}$ when $\tm_p=0$, so we expect that the solutions are proportional to $\tm_p$. When $\tm_p$ is large and
\begin{equation}
\sig_0 \neq 0\, , \qquad \sig_\alpha \neq 0 , \, \sig_0+\tm_p\, , \qquad  \sig_\alpha \neq \sig_\beta \quad \textrm{for} \quad \alpha\neq\beta \, ,  \qquad \alpha,\beta = 1,\ldots,\kd\, , \label{eq:sig-generic-2}
\end{equation}
this calculation is reliable since the chiral fields have large masses. We will show that all the vacua, as counted by the Witten index of the nonlinear sigma model, lie in a region of field space where \eqref{eq:sig-generic-2} is valid. The last $\kd$ equations in \eqref{eq:CB-twisted-eqns} can be solved for $\Sig_\alpha$ in terms of $\Sig_0$ as
\begin{eqnarray}
\Sig_\alpha & = & \frac{\Sig_0+\tm_p}{1+q_1^{1/n}\omega^{n_\alpha}} \, ,  
\end{eqnarray}
where $\omega = e^{\frac{2\pi i}{n}}$, $n_\alpha = 0,1,\ldots,n-1$. The conditions \eqref{eq:sig-generic-2} require that $n_\alpha \neq n_\beta$ for $\alpha\neq\beta$. Furthermore, when the $U(\kd)$ gauge group is broken by the diagonal \textsc{vev}s of $\Sig_{U(\kd)}$ in equation \eqref{eq:gen-sig-vevs}, there is a residual $S_{\kd}$ permutation symmetry which we can fix by requiring $0\leq n_1 < \ldots < n_\kd < n$.  The number of inequivalent solutions for the $\Sig_\alpha$ in terms of $\Sig_0$ is therefore ${n \choose \kd}$.  Upon substitution into the first equation of \eqref{eq:CB-twisted-eqns}, we obtain a degree $n \kd$ equation for $\Sig_0$ which has $n \kd = D+1$ solutions.  We thus find a total of
\begin{equation}
(D+1) {n \choose \kd}
\end{equation}
solutions, exactly matching the Witten index of the Higgs phase \eqref{eq:vac-nlsm}. This is evidence that we have not missed vacua in the nonabelian regime, where \eqref{eq:sig-generic} is not valid, and so we claim to obtain all singular loci in FI parameter space from the calculation that led to the divisors given by \eqref{eq:sing-locus-param} and \eqref{eq:sing-locus-implicit}, as well as \eqref{eq:mixed-divisor} when $\lfloor\sqrt{nk-1}\rfloor > k$ and $\kd>1$.  
 Note that there are no contributions to the calculation from the mixed branches as these are lifted in the presence of the twisted masses $\tm_p$. 
Again, since the argument employs the Witten index, we cannot rule out the possibility of bosonic/fermionic vacua that occur in pairs and contribute to the singular locus. In addition, there could be vacua corresponding to a mixed Coulomb--Higgs branch with massless nonabelian gauge bosons in the analysis of section \ref{sec:CB-mixed}; the arguments in this section cannot exclude this possibility.

\section{Examples}
\label{sec:examples}

In this section, we apply the general ideas developed in the earlier sections to specific examples.  We first focus on linear determinantal varieties of $n\times n$ matrices in projective space $\P^D$.  These are realized geometrically by the bundles $\cE=\cO^{\oplus n}$ and $\cF=\cO(1)^{\oplus n}$. According to \eqref{eq:vanChern}, the dimension will be  $D-(n-k)^2$ with vanishing first Chern class imposing $D+1-(n-k)n=0$.  Thus, in this class of examples we find three Calabi--Yau threefolds:
\begin{equation*}
(k,n) = (4,5)\, ,\  (2,4)\, ,\ (1,5)\, ,
\end{equation*}
embedded in the projective spaces $\P^4$, $\P^7$, and $\P^{19}$, respectively.  As discussed in Section~\ref{sec:det-var}, these linear determinantal varieties furnish a special class of determinantal varieties. Therefore, for our final example we analyze a determinantal Calabi--Yau threefold of a more general type.

\subsection{Resolved Determinantal Quintic in $\mathbb{P}^{4}$: $(h^{1,1},h^{2,1}) = (2,52)$} \label{sec:detQuintic}

The determinantal quintic and its small resolutions were studied in \cite{Schoen, gross2001calabi}. Hosono and Takagi analyzed its quantum moduli space in a recent paper \cite{Hosono:2011vd}, and we will see that the GLSM analysis agrees with their results. In this case, $(k,n)=(4,5)$ with PAX matter content shown in Table \ref{tab:15-GLSM}.  The incidence correspondence is
\begin{equation} \label{eq:ICExi}
X_A = \big\{ \ (\phi,x) \in \P^{4}\times \P^4 \ \big| \ (A^a\phi_a)x = 0\ \big\}\ ,
\end{equation}
where the $A^a$ are five $5\times 5$ matrices.  The determinantal variety $Z(A,4) \subset \mathbb{P}^4$ generically has isolated singular nodal points since $\codim(Z(A,3)) = 4$, so $X_A$ is a resolution of $Z(A,4)$.  Using the Thom--Porteous formula \cite{Fulton1997}, noting that $\cE=\cO^{\oplus 5}$ and $\cF=\cO(1)^{\oplus 5}$, and denoting the hyperplane class of $\P^4$ by $H$, then the cohomology class of the singular locus is
\begin{equation}
  \big[Z(A,3)\big] \,=\, \left| \begin{matrix} c_2(\cF) & c_3(\cF) \\ c_1(\cF) & c_2(\cF) \end{matrix} \right| \,=\,
     100 H^4 - 50 H^4 \,=\, 50 H^4 \ ,
\end{equation}
so we see that $Z(A,3)$ is generically composed of $50$ singular nodal points.  Using the intersection formulas \eqref{eq:Euler}-\eqref{eq:Inter}, we determine the topological data of $X_A$ to be 
\begin{equation} \label{eq:TDexi}
\begin{aligned}
    \chi(X_A)\,&=\,-100 \ , && & c_2(X_A) \cdot H \,&=\,50 \ , && &  c_2(X_A) \cdot \sigma_1 \,&=\,50 \ ,  \\
    H^3 \,&=\, 5 \ , && & H^2\cdot\sigma_1 \,&=\, 10 \ , && &  H\cdot\sigma_1^2 \,&=\, 10 \ , && & \sigma_1^3\,&=\,5 \ , 
\end{aligned}
\end{equation}
where $\sigma_1$ is the hyperplane class of the second $\P^4$ in \eqref{eq:ICExi}. On $X_A$, which has $(h^{1,1},h^{2,1})=(2,52)$, the induced classes $H$ and $\sigma_1$ are the positive generators of the K\"ahler cone $\mathcal{K}(X_A)$ as well as the integral generators of $H^2(X_A,\Z)$ .  In particular, in the phase $X_{A^T}$ the K\"ahler cone generators $H$ and $\sigma_1$ are asymptotically identified with the FI parameters $r_0+r_1$ and $r_1$, respectively.

Using mirror symmetry, Hosono and Takagi argued that the quantum K\"ahler moduli space of this Calabi--Yau has three large volume points. The discriminant locus of the mirror family can be determined using standard toric methods and is given by a curve \cite{Hosono:2011vd}
\begin{equation} \label{eq:HT-curve}
\cC: (u+v+w)^5 - 5^4uvw(u+v+w)^2+5^5 uvw(uv+vw+wu) = 0 \ \subset \ \P^2[u,v,w] \, , 
\end{equation}
in addition to three lines given by the three toric divisors
\begin{equation} \label{eq:HT-bdydivs}
 u=0 \ , \qquad v=0 \ , \qquad w=0 \ .
\end{equation}
The three large volume points are located at the intersections of two toric divisors, namely at $u=v=0$, $u=w=0$, and $v=w=0$. These predictions from mirror symmetry are in agreement with the following GLSM analysis.

\begin{table}[t]
\begin{tabular}{|c|c|c|}
\hline
	& $U(1)$ 	& $U(1)$ \\
\hline
$X_i$	&  $-1$ 	& $0$ \\
$P_i$ 	& $+1$	& $-1$ \\
$\Phi_a$	& $0$	&	$1$ \\	
\hline
\end{tabular}
\centering
\caption{Chiral superfields in the PAX GLSM for the determinantal quintic and their representations under the gauge group.  $i=1,\ldots,5$, and $a=1,\ldots,5$.} \label{tab:15-GLSM}
\end{table}

\subsubsection*{Higgs branch}

The analysis of Section \ref{sec:classical-moduli-space} shows that there are three large volume phases: $X_{A}$, $X_{A^T}$, and $Y_A$. There is no mixed Coulomb--Higgs branch separating the phases $Y_{A}$ and $Y_{A^T}$ since condition \eqref{eq:mixed-branch-condition} is not satisfied.

As argued in \cite{Hosono:2011vd}, the mutually dual incidence correspondences --- those applied to the matrix $A^a\phi_a$ and the dual matrix $(A^a\phi_a)^T$ --- relate the two large volume phases $X_A$ and $X_{A^T}$ by a flop. We can check this here explicitly as follows: as we approach the boundary of the two phases from an interior point of $X_A$, the K\"ahler parameter $-r_1>0$ (which is associated to the second $\P^4$ in \eqref{eq:ICExi}) decreases, so curves in $X_A$ with classes proportional to $\sigma_1^2$ shrink to zero size.  All curves in these classes are just the blown-up $\P^1$s (and their multi-covers) of the $50$ nodal points discussed above. These $\P^1$s are all in the same homology class, $\eta$, and as we cross into the dual phase $X_{A^T}$ with $r_1>0$, the flopped curves emerge in the homology class $\eta'$.  In the transition, intersection numbers with the divisor $D_\eta(X_A)$ dual to $\eta$ flip sign, so we should identify the divisor $D_{\eta}(X_{A^T})$ with the divisor $-D_{\eta'}(X_{A^T})$ \cite{Witten:1993yc}. Then for the described flop transition between the phases  $r_1=-\infty$ and $r_1=+\infty$, the intersections of $D_\eta(X_A)$ and $D_{\eta}(X_{A^T})$ are related by \cite{Wilson:1997} 
\begin{equation} \label{eq:flop}
\begin{aligned}   
   D_{\eta}(X_{A^T})^3 \,&=\, D_\eta(X_A)^3 - \sum_{d>0}d^3\,n_d(\eta) \ , \\ 
   D_{\eta}(X_{A^T})\cdot c_2(X_{A^T}) \,&=\, D_{\eta}(X_A)\cdot c_2(X_{A}) + 2 \sum_{d>0}d\,n_d(\eta) \ ,
\end{aligned}
\end{equation}    
where $n_d(\eta)$ are the Gromov--Witten invariants in the homology class $d\eta$. In our case, $n_1(\eta)=50$ (corresponding to the resolved $50$ nodes) and $n_d(\eta)=0$ for $d>1$ (c.f., \cite{gross2001calabi,Hosono:2011vd}), while $D_\eta(X_A)=2 H - \sigma_1$ according to \eqref{eq:TDexi}. Furthermore, since the entries of $A$ are linear, the two small resolutions $X_A$ and $X_{A^T}$ are of the same topological type. Thus, the intersections must obey 
\be
D_{\eta}(X_{A^T})^3=-D_\eta(X_A)^3 \, , \qquad D_{\eta}(X_{A^T})\cdot c_2(X_{A^T})=-D_{\eta}(X_A)\cdot c_2(X_{A}) \, .
\ee
This agrees with the relation \eqref{eq:flop} for the calculated topological data \eqref{eq:TDexi} when we associate the GLSM K\"ahler parameter $-r_1$ in $X_A$ with the $(1,1)$-form $\sigma_1$. 

Altogether, we find that the GLSM yields the three distinct large volume phases $X_A$, $X_{A^T}$ and $Y_A$, and we have seen that the two large volume phases $X_A$ and $X_{A^T}$ are related by a flop transition while remaining of the same topological type.  This is in agreement with the phase structure discovered in \cite{Hosono:2011vd}, where it was further demonstrated that the third large volume phase $Y_A$ is related to $X_A$ and $X_{A^T}$ by a similar flop transition where the roles of the two $\P^4$ factors in the incidence correspondence \eqref{eq:ICExi} are exchanged.

\subsubsection*{Discriminant locus}
The quantum complexified K\"ahler moduli space of a nonlinear sigma model contains a singular locus where correlation functions diverge; this can be interpreted as the discriminant locus of the Picard--Fuchs system of the mirror family.  Since the GLSM flows to the nonlinear sigma model, the singular locus of the GLSM should agree with the predictions of mirror symmetry (at least, up to a redefinition of the K\"ahler parameters).  This was first checked in several examples in \cite{summing}, and for GLSMs with nonabelian gauge symmetry in \cite{Hori:2006dk, Hori:2011pd}. We now check that the discriminant locus computed by Hosono and Takagi agrees with the GLSM singular locus.

The curve $\cC$ \eqref{eq:HT-curve} is a degree five curve with six nodes and therefore has a rational parameterization: it takes the form
\begin{equation}
u = \xi^5\, , \qquad v = -(\xi+1)^5\, , \label{eq:HT-curve-param}
\end{equation}
in the patch $w=1$. The Coulomb branch analysis for $(\kd,n)=(1,5)$ gives a single component discriminant locus parameterized as \eqref{eq:sing-locus-param}
\begin{equation}
(q_0(\xi),q_1(\xi)) = \left(-\frac{(1+\xi)^5}{\xi^5}, \xi^5\right)\, . \label{eq:GLSM-HT-curve}
\end{equation}
The discriminant locus \eqref{eq:GLSM-HT-curve} agrees with \eqref{eq:HT-curve} if we make the identifications
\begin{equation}
u \rightarrow q_1, \ v \rightarrow q_0q_1\, .
\end{equation}
Furthermore, we identify the large volume limit $q_1\rightarrow 0$ and $q_0q_1\rightarrow 0$ of phase $X_{A^T}$, which is associated to the K\"ahler cone generated by $H$ and $\sigma_1$, with the large volume point $u=v=0$.  Similarly, the large volume limits of the remaining phases can be associated with the points at $u=w=0$ and $v=w=0$.

This is a non-trivial test of our methods: the singular locus as predicted by the GLSM agrees with the predictions from mirror symmetry.

\subsection{Codimension 4 in $\mathbb{P}^7$: $(h^{1,1},h^{2,1})=(2,34)$}

As our second example, take $(k,n)=(2,4)$ and consider the linear determinantal Calabi--Yau threefold
\begin{equation}
    Z(A,2)\,=\, \big\{\ \phi \in \P^7\  \big| \ \rank(A^a\phi_a) \leq 2\ \big\}\  ,
\end{equation}
where the $A^a\phi_a$ is a $4\times 4$-matrix of homogeneous degree one.  Since $\codim(Z(A,1))=9$, $Z(A,2)$ is non-singular for generic choices of $A^a\phi_a$ and the variety $Z(A,2)$ is isomorphic to the incidence correspondence $X_A$.  Nevertheless, the incidence correspondence $X_A$ for $Z(A,2)$ proves useful in calculating topological data through \eqref{eq:Euler}-\eqref{eq:Inter}:
\begin{equation} \label{eq:TDexi2}
\begin{aligned}
    \chi(X_A)\,&=\,-64 \ , && & c_2(X_A) \cdot H \,&=\,56 \ , && &  c_2(X_A) \cdot \sigma_1 \,&=\,56 \ ,  \\
    H^3 \,&=\, 20 \ ,  && & H^2\cdot\sigma_1 \,&=\, 20 \ , && &  H\cdot\sigma_1^2 \,&=\, 16 \, , && & \sigma_1^3\,&=\,8 \ ,
\end{aligned}
\end{equation}
where $H$ is the hyperplane class of $\P^7$ and $\sigma_1=c_1(\cQ)$ is the Schubert class of the Grassmannian factor $G(2,4)$ in the incidence correspondence \eqref{eq:wXkfour}.  In the phase $X_{A^T}$ the hyperplane class $H$ and the Schubert class $\sigma_1$ are asymptotically identified with the FI parameters $r_0+2 r_1$ and $r_1$, respectively. The resulting Euler characteristic $\chi(X_A)=-64$ is compatible with the Hodge numbers $(h^{1,1},h^{2,1})=(2,34)$ that were derived in \cite{Kapustka0802}.  Note that Gulliksen and Neg{\aa}rd first implicitly described this particular Calabi--Yau threefold in terms of a resolution of the corresponding ideal sheaf \cite{gulliksen1971}; the variety was analyzed as a determinantal variety in \cite{Bertin0701,Kapustka0802}.

\begin{table}
\begin{tabular}{|c|c|c|}
\hline
	& $U(2)$ 	& $U(1)$ \\
\hline
$X_i$	&  $\mathbf{\bar{2}}$ 	& $0$ \\
$P_i$ & $\mathbf{{2}}$		& $-1$ \\
$\Phi_a$	& $1$		&	$1$ \\	
\hline
\end{tabular}
\centering
\caption{Chiral spectrum of the PAX model associated to the $(k,n)=(2,4)$ determinantal Calabi--Yau threefold in $\P^7$.  $i=1,\ldots,4$, and $a=1,\ldots,8$.} \label{tab:GN-GLSM}
\end{table}

We analyze the phase structure of $X_A$ using the $U(2)\times U(1)$ PAX model with matter content in Table~\ref{tab:GN-GLSM}.  The resulting phase structure of this PAX model reflects the general features discussed in detail in Section~\ref{sec:linear-det-var}, so we will be brief here. The Higgs branch yields the incidence correspondence 
\begin{equation} \label{eq:IndExii}
   X_A \,=\, \big\{ \ (\phi,x) \in \P^7\times G(2,4)\  \big| \ (A^a\phi_a)x = 0\ \big\}\  .
\end{equation}
To get a better geometric picture of the role of the two K\"ahler moduli, we can alternatively interpret this incidence correspondence as arising from desingularizing the singular variety
\begin{equation} 
Z(\mathcal A,7)\,=\, \{ \ x \in G(2,4) \ | \ \rank\,\mathcal{A} \le 7 \ \} \ ,
\end{equation}
defined in terms of the $8\times 8$ matrix  $\cA^a_{i\alpha}:=A^a_{ij}x_{j\alpha}$ that is linear in the Grassmannian planes $x \in G(2,4)$.  Since $\codim(Z(\cA,6)) = 4$ and $\dim(G(2,4))=4$, this is generically singular in points.  In this formulation, $\mathcal A$ is viewed as a section of $\Hom(\widehat \cE,\widehat \cF)$ with $\widehat\cE\cong \cO^{\oplus 8}$ and $\widehat\cF\cong\cU \otimes \cO^{\oplus 4}$, where $\cO$ and $\cU$ denote the trivial line bundle and the rank two universal subbundle of $G(2,4)$, respectively. The total Chern class of the vector bundle $\widehat\cF$ reads 
\be
c(\widehat\cF)=1-4\sigma_1+(6\sigma_1^2+4\sigma_2)-20\sigma_1\sigma_2 + \ldots \, ,
\ee
where $\sigma_1=c_1(\cQ)$ and $\sigma_2=c_2(\cQ)$ are the Schubert classes of the Grassmannians $G(2,4)$. This allows us to determine the zero-dimensional singular locus using the Thom--Porteous formula \cite{Fulton1997}:
\begin{equation}
     \big[ Z(\cA,6) \big] \,=\, \left| \begin{matrix} c_2(\widehat\cF) & c_3(\widehat\cF) \\ c_1(\widehat\cF) & c_2(\widehat\cF) \end{matrix} \right| \,=\,
     36 \sigma_1^4 - 32 \sigma_1^2 \sigma_2 + 16 \sigma_2^2 \,=\, 56 \,\{2,2\} \ .
\end{equation}
In the last step, we applied the Littlewood--Richardson rule to the Schubert classes $\sigma_k$, namely $\sigma_1^4 = 2\,\{2,2\}$ and $\sigma_1^2\sigma_2 = \sigma_2^2 = \{2,2\}$, where $\{2,2\}$ represents the cohomology class of a point in $G(2,4)$.  In particular, this shows that $Z(\cA,7)$ is generically singular at $56$ nodal points.

Thus, we can view the incidence correspondence \eqref{eq:IndExii} as a small resolution of the singular variety $Z(\mathcal A,7)$ with $56$ nodal points. In one phase of the $U(1)\times U(2)$ PAX model, the K\"ahler parameter of the $U(1)$ factor measures the volumes of the $56$ blown-up $\P^1$s with homology class $\eta$, while the other K\"ahler parameter controls the size of cycles in the singular variety $Z(\mathcal A,7)$. The other small resolution is realized in terms of the dual incidence correspondence with respect to the dual $8\times8$ matrix $\mathcal A^T$. As before, by comparing the two distinct but topologically equivalent small resolutions, we can identify the divisor $D_\eta$ dual to $\eta$ --- which is here given by $D_\eta=2\sigma_1 - H$ --- and check the flop relations \eqref{eq:flop}.  Indeed they are fulfilled for $n_1(\eta)=56$ and $n_d(\eta)=0$ for $d>1$.\footnote{Note that in order for \eqref{eq:flop} to hold, $D_\eta$ must be an integral generator of $H^2(X_A,\Z)$. Thus, with the integral structure of the intersections~\eqref{eq:TDexi2}, we can infer that $H$ and $\sigma_1$ must furnish an integral basis for $H^2(X_A,\Z)$.}

The GLSM singular locus is given by \eqref{eq:sing-locus-param} in parametric form, and by \eqref{eq:sing-locus-implicit} in implicit form.  For $k=2$, $n=4$, the resultant can easily be explicitly evaluated. We can define $\hat{q}_0 := q_0 q_1$ since $(\hat{t}_0,t_1)$ will still be an integral basis, yielding singular loci
\begin{gather}
 (1 - q_1)^4 - 2 \hat{q}_0 q_1 (1 + 6 q_1 + q_1^2)
  + \hat{q}_0^2 q_1^2= 0  \, , \nonumber \\
-(1-q_1)^8 + 4\hat{q}_0 q_1 (1-34 q_1 + q_1^2) (1-q_1)^4 - 2 \hat{q}_0^2 q_1^2 \big( 3 + 372 q_1 + 1298 q_1^2 + 372 q_1^3 + 3 q_1^4 \big)  \label{eq:sing-locus-GN} \\
+ 4 \hat{q}_0^3 q_1^3 \big(1 - 34 q_1 + q_1^2\big) - \hat{q}_0^4 q_1^4  =  0\nonumber \, .
\end{gather}
In fact, these divisors are invariant (up to multiplication by powers of $q_1$) under inversion of $q_1$.  Quotienting by this symmetry would yield simpler expressions for these divisors (while also introducing new divisors), but the connection with the GLSM parameters would be obscured so we resist.

The mirror family for this Calabi--Yau is not known. The GLSM predicts that the mirror family contains  three large complex structure points characterized by maximal unipotent monodromy, and it has a discriminant locus consisting of the two rational curves in \eqref{eq:sing-locus-GN}, and boundary divisors associated to the FI parameters tending to infinity.

\subsection{Codimension 16 in $\mathbb{P}^{19}$: $(h^{1,1},h^{2,1})=(2,52)$}

This Calabi--Yau, with $(k,n)=(1,5)$, is defined by the incidence correspondence
\begin{equation}
X_A = \big\{\ (\phi,x) \in \P^{19}\times G(4,5)\ \big| \ (A^a\phi_a) x = 0\ \big\}\  , \label{eq:def-45-example}
\end{equation}
and is actually isomorphic to the resolved determinantal quintic.  Consider the vector space of $5\times 5$ complex matrices with the hermitian inner product $\langle A,B \rangle := \tr (A^\dagger B)$. The matrices $A^a$ span a 20-dimensional subspace; let $B^b$, $b=1,\ldots,5$, be generic matrices from the orthogonal subspace with $\tr(A^a B^b)=0$. The defining equation \eqref{eq:def-45-example} can then be rewritten as
\begin{equation}
\tr( N B^b) = 0\, , \qquad Nx=0\, , \label{eq:15-45-iso}
\end{equation}
where $N$ is an arbitrary $5\times 5$ matrix. The second condition can be solved by
\begin{equation}
N_{ij} = \psi_i \hat{x}_j\, ,
\end{equation}
where $\hat{x}_j$ are the five $4\times 4$ minors of $x$ that span its kernel and $\psi_i$ is an arbitrary vector. The $\hat{x}_j$ can also be viewed as defining the Pl\"ucker embedding, which is an isomorphism in this case, of $G(4,5)$ into $\P^4$.  The first condition in \eqref{eq:15-45-iso} now becomes $B^b_{ji}\psi_i \hat{x}_j=0$, which is the incidence correspondence corresponding to the determinantal quintic analyzed earlier.

The isomorphism to the determinantal quintic can also be confirmed by comparing intersection numbers.  For the variety $X_A$ in $\P^{19}$, the intersection calculation yields
\begin{equation} \label{eq:TDexiii}
\begin{aligned}
    \chi(X_A)\,&=\,-100 \ , && & c_2(X_A) \cdot H \,&=\,100 \ , && &  c_2(X_A) \cdot \sigma_1 \,&=\,50 \ ,  \\
    H^3 \,&=\, 70 \ ,  && & H^2\cdot\sigma_1 \,&=\, 35 \ , && &  H\cdot\sigma_1^2 \,&=\, 15 \ , && & \sigma_1^3\,&=\,5 \ ,
\end{aligned}
\end{equation}
where $H$ is the hyperplane class of $\P^{19}$ and $\sigma_1$ is the Schubert class of $G(4,5)$.  Here, $H-\sigma_1$ and $\sigma_1$ are the two generators of the K\"ahler cone $\mathcal K(X_A)$ and are readily identified with the generators of the K\"ahler cone of the resolved determinantal quintic discussed in Section~\ref{sec:detQuintic}. Here, in the phase $X_{A^T}$ the cohomology elements $H$ and $\sigma_1$ are asymptotically identified with $r_0+4 r_1$ and $r_1$, while the generators $H-\sigma_1$ and $\sigma_1$ of the K\"ahler cone are asympotically associated to $r_0+3 r_1$ and $r_1$.

We expect the singular locus of the GLSM describing this Calabi--Yau threefold to agree with the singular locus for the determinantal quintic, but we immediately encounter a puzzle: when $(k,n)=(1,5)$ the condition \eqref{eq:mixed-branch-condition} is satisfied and we have a mixed Coulomb--Higgs branch, while there was no such branch for the determinantal quintic. The discriminant locus in this example consists of two components,
\begin{equation}
q_1 + 1 = 0\, ,
\end{equation}
associated with a mixed Coulomb--Higgs branch, and one associated with a pure Coulomb branch that can be written in parametric form as
\begin{equation}
q_1 = \xi^5\, , \qquad q_0 q_1^4 = \prod_{\alpha=0}^3 (1 + \omega^\alpha \xi)^5 \, , \qquad \mbox{where } \ \ \omega = e^{2\pi i/5}\, . \label{eq:(1,5)-curve}
\end{equation}
Note that this parametric locus can be rewritten as
\begin{equation}
q_1 = \xi^5\, , \qquad \frac{(1+q_1)^5}{q_0q_1^4} = (1+\xi)^5 \, .
\end{equation}
This is precisely the parametrization of the singular locus in the determinantal quintic \eqref{eq:HT-curve-param}. The resolution of the apparent paradox is now clear: the map between the parameter spaces of the $(k,n)=(1,5)$ PAX model and the $(k,n)=(4,5)$ PAX model for the determinantal quintic takes the explicit form
\begin{equation}
u \rightarrow -\frac{(1+q_1)^5}{q_0 q_1^4}\, , \qquad v \rightarrow q_1\, .
\end{equation}
The boundary divisor $u=0$ in the moduli space of the determinantal quintic (see equation \eqref{eq:HT-bdydivs}), as determined in \cite{Hosono:2011vd}, maps to the mixed branch locus $q_1+1=0$. Moreover, the quintic curve \eqref{eq:HT-curve} maps to the Coulomb branch locus above in equation \eqref{eq:(1,5)-curve}. The singular loci in the K\"ahler moduli space of this example, therefore, are in agreement with those of the determinantal quintic.  Note that the large volume limit $q_1 \rightarrow 0$, $q_0 q_1^3\rightarrow 0$, is mapped to the large volume point $v=w=0$ (in terms of the homogeneous coordinates of $\P^2$ used in \eqref{eq:HT-curve}).

\subsection{Bertin Calabi--Yau threefolds in $\mathbb{P}^7$: $(h^{1,1},h^{2,1})=(2,58)$}

In this section, we study a determinantal Calabi--Yau threefold $Z(A,1) \subset \P^7$ defined by the vanishing of the maximal minors of a generic section $A\in \mathrm{Hom}(\cE,\cF)$, where $\mathcal{E} = \cO^{\oplus 3}$ and $\mathcal{F} = \cO(1)^{\oplus 2}\oplus \cO(2)$. This Calabi--Yau was described by Bertin in \cite{Bertin0701}.\footnote{Note that this choice of bundles corrects a typo in \cite{Bertin0701}.} 
The first Chern class of $Z(A,1)$ vanishes, as is clear from equation  \eqref{eq:vanChern}. Since the variety $Z(A,1)$ is smooth for generic choices of $A$, the incidence correspondence~\eqref{eq:wXkfour} identifies $Z(A,1)$ isomorphically with $X_A$. The Hodge numbers of $X_A$ are $h^{1,1}=2$ and $h^{2,1}=58$ \cite{Bertin0701}, and as before we extract the topological data
\begin{equation} \label{eq:TDexiv}
\begin{aligned}
    \chi(X_A)\,&=\,-112 \ , && & c_2(X_A) \cdot H \,&=\,62 \ , && &  c_2(X_A) \cdot \sigma_1 \,&=\,36 \ ,  \\
    H^3 \,&=\, 17 \ , && & \sigma_1^3\,&=\,0 \ , && & H^2\cdot\sigma_1 \,&=\,10  \ , && &  H\cdot\sigma_1^2 \,&=\,4 \ ,
\end{aligned}
\end{equation}
with the hyperplane classes $H$ of $\P^7$ and $\sigma_1$ of $\P^2\cong G(1,3)$ of the ambient space $\P^7\times\P^2$ of $X_A$.

The PAX model for this Calabi--Yau has a $U(2)\times U(1)$ gauge group, while the PAXY model has an abelian $U(1)\times U(1)$ gauge group.  We will study this Calabi--Yau with the PAXY model to avoid nonabelian dynamics.  The gauge group and matter content for the PAXY model are described in Table \ref{tab:Bertin-GLSM} and superpotential in \eqref{eqn:nonabel-spot}.
\begin{table}[t]
\begin{tabular}{|c|c|c|c|c|c|c|c|}
\hline
& $\tX_i$ & $\tY_{1,2}$ & $\tY_3$ & $\tP_{i1}$ & $\tP_{i2}$ & $\tP_{i3}$ & $\Phi_a$  \\
\hline
$U(1)_0$ & 0 & 1 & 2 & $-1$ & $-1$ & $-2$ & 1 \\
\hline
$U(1)_1$ & 1 & $-1$ & $-1$ & 0 & 0 & 0 & 0 \\
\hline
\end{tabular}
\centering
\caption{Chiral superfields in the Bertin GLSM and their representations under the gauge group.  $i=1,2,3$, and $a=1,\ldots,8$.} \label{tab:Bertin-GLSM}
\end{table}
The charge assignments yield a non-anomalous axial $U(1)$ R-symmetry and central charge $\frac{c}{3}=3$.  The D-terms are
\begin{eqnarray}
\begin{array}{rc}
U(1)_0: & |\ty_1|^2 + |\ty_2|^2 + 2|\ty_3|^2 + \sum_a |\phi_a|^2 - \sum_i \big( |\tp_{i1}|^2 + |\tp_{i2}|^2 + 2|\tp_{i3}|^2 \big) = r_0 \, , \\
U(1)_1: & \sum_i \big(|\tx_i|^2 - |\ty_i|^2\big) = r_1  \, ,
\end{array} \label{eq:Bertin-D}
\end{eqnarray}
while the F-terms read
\begin{eqnarray}
\tp \ty = 0\, , \qquad \tx  \tp = 0\, , \qquad \tr\big(\tp\, \p_a A(\phi)\big) = 0\, , \qquad A(\phi) - \ty\tx = 0\, . \label{eq:Bertin-F}
\end{eqnarray}

\subsubsection*{Classical Higgs Branch}
We first study the classical phase structure of the theory as a function of the FI parameters $(r_0,r_1)$ and then incorporate the effects of quantum corrections, as we did in Section \ref{sec:linear-det-var}. The arguments we employ are simple variants of the ones in Section \ref{sec:linear-det-var}, so we will be concise.

\noindent
\underline{\bf $r_0>0,$ $r_0+2r_1>0$}:\\
Since a generically chosen $A(\phi)$ has rank at least one for all $\phi\in V$, the D-terms and F-terms in \eqref{eq:Bertin-D}-\eqref{eq:Bertin-F} imply that $\tx$, $\ty$, and $\phi$, are all non-zero.  Furthermore,  since $Z(A,1)$ will be nonsingular, we must have $\tp=0$.  The vacuum moduli space is then 
\begin{equation}\label{eq: Bertin geom}
  \hat X_A = \big\{\ (\phi,\tx,\ty)\in \cV_{12} \ \big|  \ A(\phi) = \ty\tx \ \big\}\, ,
\end{equation}
where $(\phi,\tx,\ty)$ are homogeneous coordinates for the 12-dimensional toric variety $\cV_{12}$ defined by the D-terms \eqref{eq:Bertin-D}.  The variety $\hat X_A$, while also isomorphic to $Z(A,1)$, can be viewed as a fibration over the determinantal locus $Z(A,1)$: for any point $\phi \in Z(A,1)$, $\rank(A(\phi)) = 1$ and the fiber $(\tx,\ty)$ is uniquely determined.  
While $Z(A,1)$ is not a complete intersection in $\P^7$, the variety $\hat X_A$ is a complete intersection of the nine equations $A(\phi)_{ij} - \ty_i \tx_j = 0$ in the ambient toric variety $\cV_{12}$. From the viewpoint of the GLSM, this had to be the case since the gauge theory is abelian.

\noindent
\underline{\bf $r_1<0,$ $r_1+r_0<0$}:\\
The D-terms require that $\ty$ and $\tp$ be non-zero. If $\phi \neq 0$, the argument from the previous phase applies, and the F-terms require that $\tp$ vanish.\footnote{Schematically, the F-terms in the previous phase can be written as $A(\phi) = yx$ and $\tp\cdot \frac{\partial(A(\phi)-yx)}{\partial (\phi,x,y)}=0$. If $\phi \neq 0$, then it corresponds to a point on $\P^7$, and the smoothness of $Z(A,1)$ at that point would imply that $\tp=0$.} Since this is in conflict with the D-terms it must be the case that $\phi=0$, which in turn implies that $\tx=0$. The F-terms further impose the conditions $\tp_{i1}=0$, $\tp_{i2}=0$, and $\ty_{3}=0$.  Redefining the $\C^*$ action, we see that the moduli space appears to be the product  $\mathbb{P}^{1}\times\mathbb{P}^{2}$, with the D-terms
\begin{eqnarray}
2\sum_i |\tp_{i3}|^2 = -r_0-r_1 \, , \nonumber \\
|\ty_1|^2 + |\ty_2|^2= -r_1\, .
\end{eqnarray}
The fields $\tx$ and $\phi$ that correspond to the transverse directions cannot be integrated out since they are actually massless along a sub-locus of $\P^1\times \P^2$ but are obstructed at higher order. This phase, therefore, is a hybrid phase: a combination of a nonlinear sigma model along certain directions and a Landau--Ginzburg theory along others \cite{Witten:1993yc}. The existence of these additional massless directions is easily established by studying the superpotential for quadratic fluctuations about the vacuum
\begin{eqnarray}
W_{\rm{quad}} & = & A_{1i}^a \ \delta\tP_{i1}\, \delta\Phi_a +  A_{2i}^a \ \delta\tP_{i2}\, \delta\Phi_a + A_{3i}^{ab} \vev{\tP_{i3}}\ \delta\Phi_a \, \delta\Phi_b - \vev{\tY_1} \ \delta \tX_i \, \delta \tP_{i1} \nonumber \\
& & - \vev{\tY_2} \ \delta \tX_i \, \delta \tP_{i2} - \vev{\tP_{i3}}\ \delta \tX_i \, \delta \tY_3\, .
\end{eqnarray}
We can assemble the fluctuations into a single 18-dimensional vector,
\begin{equation}
\delta \Psi_A = \{ \delta \Phi_{a}, \ \delta \tP_{i1}, \ \delta \tP_{i2}, \ \delta \tX_i, \ \delta \tY_3\}\, ,
\end{equation}
in terms of which, we have $W_{\rm{quad}} = \delta \Psi_A \ \mathbf{M}_{AB} \ \delta \Psi_B$, with an $18\times 18$ mass matrix
\begin{equation}
\mathbf{M} = \begin{pmatrix}
\sum_i A_{3i}^{ab} \vev{\tP_{i3}} & A_{1i}^a & A_{2i}^a & 0 & 0 \\
A_{1i}^a	&	0	& 	0	& -\vev{\tY_1}	&	0 \\
A_{2i}^a	&	0	&	0	&-\vev{\tY_2}	&	0 \\
0	&-\vev{\tY_1}	&-\vev{\tY_2}	&	0	& -\vev{\tP_{i3}} \\
0	&	0	&	0	&-\vev{\tP_{i3}}	&	0
\end{pmatrix}\ .
\end{equation}
On the sub-locus of $\P^1 \times \P^2$, where $\det(\mathbf{M}) = 0$, there are additional massless directions that correspond to fluctuations of a Landau--Ginzburg model, resulting in a hybrid phase.\footnote{We would like to thank the authors of \cite{Caldararu:2007tc}, especially E.~Sharpe, for correcting an error in the analysis of this phase in an earlier version of this paper.}

\noindent
\underline{\bf $r_0<0,$ $r_1>0$}:\\
This phase is quite similar to the previous one.  In this phase, $\tx$ and $\tp$ are required to be non-zero by the D-terms.  Since $\tp$ is nonzero, the smoothness condition that was necessary in the determinantal phase implies that $\phi = 0$, which also implies $\ty=0$.    The equation $\tr\big(\tp\, \p_a A(\phi)\big) = 0$ then reduces to
\begin{equation}
\sum_{j}\tp_{j1}A^{a}_{1j}+\tp_{j2}A^{a}_{2j}=0\, .
\end{equation}
For a generic choice of the parameters $A_{ij}^{a}$ these equations imply $\tp_{j1}=\tp_{j2}=0$. The vacuum moduli space is again a hybrid like the previous phase: a nonlinear sigma model with target space
\begin{equation}
\big\{\ (\tx_j,\tp_{j3}) \in \P^2 \times \P^2\ \big| \ \sum_{j}\tx_{j}\tp_{j3}=0 \ \big\}\, ,
\end{equation}
and a Landau--Ginzburg model along the null directions of the corresponding mass matrix.

\noindent
\underline{\bf $r_0+2r_1<0,$ $r_1+r_0>0$}:\\
In this phase, either $\ty$ or $\tp$ is non-zero; the moduli space has different branches of solutions, resulting in a mixture of the phases analyzed above.

\subsubsection*{Coulomb Branch}
Along the Coulomb branch, where $\sigma$-fields can have large \textsc{vev}s, we integrate out the massive chiral fields to obtain the effective twisted superpotential for $\sigma$.  Using the formula for the one-loop correction found in \cite{Witten:1993yc, summing}, we obtain
\begin{eqnarray}
2\sqrt{2}\, \widetilde{W}_{\mathit{eff}}(\Sigma)=& & \Sigma_{0}\left(it_{0}-\frac{1}{2\pi}\log\left(\frac{(\Sigma_{0}-\Sigma_{1})^{2}(2\Sigma_{0}-\Sigma_{1})^{2}}{64\Sigma_{0}^{4}}\right)\right)\nonumber \\
&  & +\Sigma_{1}\left(it_{1} + \frac{1}{2\pi}\log\left(\frac{(\Sigma_{0}-\Sigma_{1})^{2}(2\Sigma_{0}-\Sigma_{1})}{\Sigma_{1}^{3}}\right)\right)\, .
\end{eqnarray}
Critical points exist only when the FI parameters lie along the curve 
\begin{eqnarray}
\big(q_{0}(\xi),q_{1}(\xi)
\big)=\left(\frac{\xi^{2}(1+\xi)^{2}}{64},\frac{(1-\xi)^{3}}{\xi^{2}(1+\xi)}\right)\, ,
\end{eqnarray}
where $q_{i} = e^{2\pi i t_{i}}$ and $\xi = \frac{\Sigma_0 - \Sigma_1}{\Sigma_0}$. The parameterization defines the rational, quartic curve
\begin{gather}
-1+48 q_0-768 q_0^2+4096 q_0^3+136 q_0 q_1+10624 q_0^2 q_1+16384 q_0^3 q_1+q_0 q_1^2-5120 q_0^2 q_1^2  \nonumber \\
+24576 q_0^3 q_1^2-128 q_0^2 q_1^3+16384 q_0^3 q_1^3+4096 q_0^3 q_1^4 = 0 \, ,
\end{gather}
yielding a prediction to be tested against future work.

\section{Conclusions and Future Directions}
In this paper, we studied the PAX and PAXY models, describing two classes of nonabelian supersymmetric GLSMs whose vacuum moduli spaces correspond to desingularized determinantal varieties in their geometric phases.  These generalized the symplectic and orthogonal group nonabelian GLSMs studied in \cite{Hori:2006dk, Hori:2011pd} to unitary gauge groups, and expanded GLSM methods to a larger class of examples than those provided by complete intersection Calabi--Yau varieties.  (In fact, the latter may be viewed as trivial determinantal varieties.)

We then argued that the PAX and PAXY models yield the same low energy physics, realizing a two-dimensional version of Seiberg duality recently explained by Hori in \cite{Hori:2011pd}, by showing that their geometric phases give rise to distinct but isomorphic incidence correspondences that desingularize the same determinantal variety.  In particular, the incidence correspondence associated to the PAXY model realizes the desingularized determinantal variety as the zero locus of a global section of a vector bundle over a compact space, which in turn is itself a Grassmannian fibration over the ambient toric variety. Thus, the PAXY model realizes determinantal Calabi--Yau varieties as complete intersections in higher dimensional (and in general non-toric) embedding spaces.\footnote{Tj{\o}tta employed a similar construction \cite{Tjotta1999}, in order to describe the Pfaffian Calabi--Yau threefold studied by R{\o}dland \cite{Rodland9801} as the zero locus of a global section of a vector bundle.}  
To analyze the moduli spaces and topological invariants of these determinantal Calabi--Yau varieties, we then employed techniques complementary to the homological algebra approach \cite{Burch, gulliksen1971, buchsbaum1977algebra, kustinmiller, Decker:1994, Tonoli2006, Bertin0701, Kapustka0802}. 

Using techniques developed in \cite{summing} for abelian GLSMs and further extended in \cite{Hori:2006dk} for nonabelian GLSMs, we then analyzed the phase structure of the quantum-corrected K\"ahler moduli spaces of linear determinantal varieties in projective space. In particular, we identified the singular loci in the quantum K\"ahler moduli space where non-compact Coulomb branches appear in the GLSM, corresponding to singularities in correlation functions of the IR superconformal field theory.  We explicitly computed the singular locus in parametric form, finding that it consists of a union of irreducible rational curves and explaining how a similar computation of the quantum K\"ahler moduli space can be carried out for any determinantal subvariety of a compact toric variety. We also presented evidence against additional components of the discriminant locus corresponding to nonabelian dynamics.

We next applied these GLSM methods to a few examples of determinantal Calabi--Yau varieties.  The first example concerned the quintic determinantal variety in $\P^4$, studied in detail in \cite{Hosono:2011vd}, providing a check of our techniques: the singular loci of the quantum K\"ahler moduli space computed by the GLSM agrees with the predictions from mirror symmetry. 
Furthermore, using the PAXY incidence correspondence we determined certain topological invariants, finding agreement with the previously known results.  Finally, we repeated the GLSM analysis for other examples of determinantal Calabi--Yau varieties, finding the singularities in their quantum K\"ahler moduli spaces and computing their topological invariants near the large volume point. To our knowledge, the mirror families of the latter examples are not known, so we hope these data will prove useful in identifying the mirror families of these determinantal Calabi--Yau varieties.

Mirror symmetry for complete intersections in toric varieties is understood through the duality of polytopes \cite{Batyrev1993, borisov-ms, Batyrev-Borisov-1994}, which also led to a mirror construction for complete intersection Calabi--Yau manifolds in Grassmannians \cite{batyrev1998conifold} and more general partial flag manifolds \cite{batyrev2000mirror}.  For more general Calabi--Yau varieties it is not known how to find their mirrors, but an interesting proposal has been put forward in \cite{arXiv:0708.4402,arXiv:1103.2673}.  In abelian GLSMs describing complete intersections in toric varieties, mirror symmetry can be interpreted as abelian particle-vortex duality \cite{towards-duality, Hori:2000kt, Hori:2003ic}.  A nonabelian generalization of this story, possibly along the lines proposed in \cite{Hori:2000kt} or in \cite{arXiv:0708.4402,arXiv:1103.2673}, could thus shed light on mirror symmetry for non-complete intersection Calabi--Yau varieties.\footnote{Other useful references concerning mirror symmetry of non-complete intersection Calabi--Yau varieties include \cite{Tjotta1999,hosonokonishi0704,arXiv:1011.2350}.}  A check on such a proposal could be performed by comparison with mirror constructions in \cite{kanazawa2010pfaffian} for certain Pfaffian Calabi--Yau threefolds that appear in \cite{Tonoli2006}.  We leave this to future work. 

Another direction to pursue would be to generalize this construction to determinantal varieties defined by non-square matrices.  As we discussed in Section \ref{sec:generalize}, the axial $U(1)$ R-symmetry in the PAX and PAXY models is anomalous in this case since the $U(k)$ gauge theory has unequal numbers of fundamentals and anti-fundamentals.  Since we are not aware of any examples of non-square determinantal varieties (without symmetry properties) that are Calabi--Yau, it would be useful to construct explicit examples to help guide efforts to construct a corresponding GLSM.

We should emphasize that determinantal varieties form only a special subset of non-complete intersection Calabi--Yau varieties. One expects that a non-complete intersection of high codimension has an ideal sheaf with  a complicated resolution. In particular, this means that the equations have non-trivial relations and higher order syzygies. In the determinantal case, we have successfully encoded the syzygies into a GLSM superpotential.  It would be interesting to find similar constructions for other types of Calabi--Yau ideals.

We hope to return to some of these questions in a sequel.


\bigskip
\subsection*{Acknowledgments}

We would like to thank
Janko B\"ohm,
Xenia de la Ossa,
Ron Donagi,
Mike Douglas,
Tom Faulkner,
Atsushi Kanazawa,
Sheldon Katz,
Kentaro Hori,
Shinobu Hosono,
Nabil Iqbal,
Albrecht Klemm,
Johanna Knapp,
Ronen Plesser,
Eric Sharpe,
and Johannes Walcher
for useful discussions and correspondence. 
We particularly thank Kentaro Hori and Johanna Knapp for communicating
their work in progress to us.
We also thank Eric Sharpe and the other authors of \cite{Caldararu:2007tc} for pointing out an error in the analysis in section 5.4 in the first version of this paper.
We thank Kentaro Hori for pointing out an insufficiency in the analysis of the phase structure in the second version.
H.J., J.M.L. and D.R.M. would like to thank the Banff International Research Station,
and H.J., V.K., D.R.M. and M.R. the Simons Center for Geometry and Physics for hospitality at various stages of this project.
H.J. and J.M.L. would also like to thank the Kavli Institute for Theoretical Physics, where this project was initiated; D.R.M. would also like to thank the Aspen Center for Physics for hospitality.
H.J. is supported by the DFG grant KL 2271/1-1;
V.K. is supported in part by the National Science Foundation under Grant No. PHY11-25915;
J.M.L. is supported by the National Science and Engineering Research Council of Canada;
D.R.M. is supported in part by NSF Grants DMS-1007414 and PHY-1066293.
\bigskip
\goodbreak

\ifx\undefined\bysame
\newcommand{\bysame}{\leavevmode\hbox to3em{\hrulefill}\,}
\fi


\begin{thebibliography}{10}

\bibitem{Witten:1993yc}
E.~Witten, {\em Phases of {$N = 2$} theories in two dimensions}, Nucl. Phys. B
  {\bf 403} (1993) 159--222, {\tt arXiv:hep-th/9301042}.

\bibitem{summing}
D.~R. Morrison and M.~R. Plesser, {\em Summing the instantons: Quantum
  cohomology and mirror symmetry in toric varieties}, Nuclear Phys. B {\bf 440}
  (1995) 279--354, {\tt arXiv:hep-th/9412236}.

\bibitem{serre-codim-2}
J.-P. Serre, {\em Sur les modules projectifs}, Alg{\`e}bre et th{\'e}orie des
  nombres, S{\'e}minaire Dubreil, vol.~14, Secr{\'e}tariat math{\'e}matique,
  Paris, 1960-1961.

\bibitem{buchsbaum1977algebra}
D.~A. Buchsbaum and D.~Eisenbud, {\em Algebra structures for finite free
  resolutions, and some structure theorems for ideals of codimension {$3$}},
  Amer. J. Math. {\bf 99} (1977) 447--485.

\bibitem{okonek1994notes}
C.~Okonek, {\em Notes on varieties of codimension {$3$} in {$\mathbb{P}^N$}},
  Manuscripta Math. {\bf 84} (1994) 421--442.

\bibitem{walter1996pfaffian}
C.~H. Walter, {\em Pfaffian subschemes}, J. Algebraic Geom. {\bf 5} (1996)
  671--704, {\tt arXiv:alg-geom/9406005}.

\bibitem{Tonoli2006}
F.~Tonoli, {\em Construction of {C}alabi--{Y}au 3-folds in {$\mathbb P^6$}}, J.
  Algebraic Geom. {\bf 13} (2004) 209--232.

\bibitem{kustinmiller}
A.~Kustin and M.~Miller, {\em Structure theory for a class of grade four
  {G}orenstein ideals}, Trans. Amer. Math. Soc. {\bf 270} (1982) 287--307.

\bibitem{gulliksen1971}
T.~H. Gulliksen and O.~G. Neg{\aa}rd, {\em Un complexe r\'esolvant pour
  certains id\'eaux d\'eterminantiels}, C. R. Acad. Sci. Paris S\'er. A-B {\bf
  274} (1972) A16--A18.

\bibitem{Harris1995}
J.~Harris, {\em Algebraic geometry}, Graduate Texts in Mathematics, vol. 133,
  Springer-Verlag, New York, 1992.

\bibitem{Hori:2006dk}
K.~Hori and D.~Tong, {\em Aspects of non-abelian gauge dynamics in
  two-dimensional {$\mathcal{N}=(2,2)$} theories}, J. High Energy Phys. (2007)
  079, {\tt arXiv:hep-th/0609032}.

\bibitem{Rodland9801}
E.~A. R{\o}dland, {\em The {P}faffian {C}alabi--{Y}au, its mirror, and their
  link to the {G}rassmannian {$G(2,7)$}}, Compositio Math. {\bf 122} (2000)
  135--149, {\tt arXiv:math.AG/9801092}.

\bibitem{Witten:1993xi}
E.~Witten, {\em The {V}erlinde algebra and the cohomology of the
  {G}rassmannian}, Geometry, Topology, {\&} Physics for Raoul Bott (S.-T. Yau,
  ed.), Conf. Proc. Lecture Notes Geom. Topology, vol.~IV, Int. Press,
  Cambridge, MA, 1995, pp.~357--422, {\tt arXiv:hep-th/9312104}.

\bibitem{Lerche:2001vj}
W.~Lerche, P.~Mayr, and J.~Walcher, {\em {A New kind of McKay correspondence
  from nonAbelian gauge theories}}, 2001, {\tt arXiv:hep-th/0103114} {\tt
  [hep-th]}.

\bibitem{Donagi:2007hi}
R.~Donagi and E.~Sharpe, {\em G{LSM}s for partial flag manifolds}, J. Geom.
  Phys. {\bf 58} (2008) 1662--1692, {\tt arXiv:0704.1761 [hep-th]}.

\bibitem{Hori:2011pd}
K.~Hori, {\em Duality in two-dimensional {(2,2)} supersymmetric non-{A}belian
  gauge theories}, {\tt arXiv:1104.2853 [hep-th]}.

\bibitem{HoriKnappLectures}
K.~Hori and J.~Knapp, Lectures at the Workshop on Noncommutative Algebraic
  Geometry and D-branes, Simons Center for Geometry and Physics, December 2011.

\bibitem{Bertin0701}
M.-A. Bertin, {\em Examples of {C}alabi--{Y}au 3-folds of {$\mathbb P^7$} with
  {$\rho=1$}}, Canad. J. Math. {\bf 61} (2009) 1050--1072, {\tt
  arXiv:math.AG/0701511}.

\bibitem{Kapustka0802}
M.~Kapustka and G.~Kapustka, {\em A cascade of determinantal {C}alabi--{Y}au
  threefolds}, Math. Nachr. {\bf 283} (2010) 1795--1809, {\tt arXiv:0802.3669
  [math.AG]}.

\bibitem{Hosono:2011vd}
S.~Hosono and H.~Takagi, {\em Mirror symmetry and projective geometry of {R}eye
  congruences {I}}, {\tt arXiv:1101.2746 [math.AG]}.

\bibitem{Seiberg:1994pq}
N.~Seiberg, {\em Electric-magnetic duality in supersymmetric nonabelian gauge
  theories}, Nucl. Phys. B {\bf 435} (1995) 129--146, {\tt
  arXiv:hep-th/9411149}.

\bibitem{Horrocks1973}
G.~Horrocks and D.~Mumford, {\em A rank {$2$} vector bundle on {${\bf P}^{4}$}
  with {$15,000$}\ symmetries}, Topology {\bf 12} (1973) 63--81.

\bibitem{Schoen}
C.~Schoen, {\em On the geometry of a special determinantal hypersurface
  associated to the {M}umford-{H}orrocks vector bundle}, J. Reine Angew. Math.
  {\bf 364} (1986) 85--111.

\bibitem{gross2001calabi}
M.~Gross and S.~Popescu, {\em Calabi--{Y}au threefolds and moduli of abelian
  surfaces. {I}}, Compositio Math. {\bf 127} (2001) 169--228, {\tt
  arXiv:math.AG/0001089}.

\bibitem{Fulton1997}
W.~Fulton, {\em Intersection theory}, Ergeb. Math. Grenzgeb. (3), vol.~2,
  Springer-Verlag, Berlin, 1984.

\bibitem{kanazawa2010pfaffian}
A.~Kanazawa, {\em On {P}faffian {C}alabi--{Y}au varieties and mirror symmetry},
  {\tt arXiv:1006.0223 [math.AG]}.

\bibitem{Tjotta1999}
E.~N. Tj{\o}tta, {\em Quantum cohomology of a {P}faffian {C}alabi--{Y}au
  variety: verifying mirror symmetry predictions}, Compositio Math. {\bf 126}
  (2001) 79--89, {\tt arXiv:math.AG/9906119}.

\bibitem{Wall1966}
C.~T.~C. Wall, {\em Classification problems in differential topology. {V}. {O}n
  certain {$6$}-manifolds}, Invent. Math. 1 (1966), 355-374; corrigendum, ibid
  {\bf 2} (1966) 306.

\bibitem{Berenstein:2002fi}
D.~Berenstein and M.~R. Douglas, {\em Seiberg duality for quiver gauge
  theories}, {\tt arXiv:hep-th/0207027}.

\bibitem{Burch}
L.~Burch, {\em On ideals of finite homological dimension in local rings}, Proc.
  Cambridge Philos. Soc. {\bf 64} (1968) 941--948.

\bibitem{Decker:1994}
W.~Decker and S.~Popescu, {\em On surfaces in {${\bf P}^4$} and {$3$}-folds in
  {${\bf P}^5$}}, Vector bundles in algebraic geometry ({D}urham, 1993), London
  Math. Soc. Lecture Note Ser., vol. 208, Cambridge Univ. Press, Cambridge,
  1995, pp.~69--100, {\tt arXiv:alg-geom/9402006}.

\bibitem{Rational-Curves-book}
J.~R. Sendra, F.~Winkler, and S.~P{\'e}rez-D{\'{\i}}az, {\em Rational algebraic
  curves: A computer algebra approach}, Algorithms and Computation in
  Mathematics, vol.~22, Springer, Berlin, 2008.

\bibitem{hilbert1993theory}
D.~Hilbert, {\em Theory of algebraic invariants}, Cambridge University Press,
  Cambridge, 1993, (Translated from the German and with a preface by Reinhard
  C. Laubenbacher, Edited and with an introduction by Bernd Sturmfels).

\bibitem{Witten:1982df}
E.~Witten, {\em Constraints on supersymmetry breaking}, Nuclear Phys. B {\bf
  202} (1982) 253--316.

\bibitem{Wilson:1997}
P.~M.~H. Wilson, {\em Flops, {T}ype {III} contractions and {G}romov--{W}itten
  invariants on {C}alabi--{Y}au threefolds}, New trends in algebraic geometry
  ({W}arwick, 1996), London Math. Soc. Lecture Note Ser., vol. 264, Cambridge
  Univ. Press, Cambridge, 1999, pp.~465--484, {\tt arXiv:alg-geom/9707008}.

\bibitem{Caldararu:2007tc}
A.~C{\u{a}}ld{\u{a}}raru, J.~Distler, S.~Hellerman, T.~Pantev, and E.~Sharpe,
  {\em Non-birational twisted derived equivalences in abelian {GLSM}s}, Comm.
  Math. Phys. {\bf 294} (2010) 605--645, {\tt arXiv:0709.3855 [hep-th]}.

\bibitem{Batyrev1993}
V.~V. Batyrev, {\em Dual polyhedra and mirror symmetry for {C}alabi--{Y}au
  hypersurfaces in toric varieties}, J. Algebraic Geom. {\bf 3} (1994)
  493--535, {\tt arXiv:alg-geom/9310003}.

\bibitem{borisov-ms}
L.~A. Borisov, {\em Towards the mirror symmetry for {C}alabi--{Y}au complete
  intersections in {G}orenstein toric {F}ano varieties}, {\tt
  arXiv:alg-geom/9310001}.

\bibitem{Batyrev-Borisov-1994}
V.~V. Batyrev and L.~A. Borisov, {\em On {C}alabi--{Y}au complete intersections
  in toric varieties}, Higher-dimensional complex varieties (Trento, 1994), de
  Gruyter, Berlin, 1996, pp.~39--65, {\tt arXiv:alg-geom/9412017}.

\bibitem{batyrev1998conifold}
V.~V. Batyrev, I.~Ciocan-Fontanine, B.~Kim, and D.~van Straten, {\em Conifold
  transitions and mirror symmetry for {C}alabi--{Y}au complete intersections in
  {G}rassmannians}, Nucl.Phys. B {\bf 514} (1998) 640--666, {\tt
  arXiv:alg-geom/9710022}.

\bibitem{batyrev2000mirror}
\bysame, {\em Mirror symmetry and toric degenerations of partial flag
  manifolds}, Acta Math. {\bf 184} (2000) 1--39, {\tt arXiv:math.AG/9803108}.

\bibitem{arXiv:0708.4402}
J.~B{\"o}hm, {\em Mirror symmetry and tropical geometry}, {\tt arXiv:0708.4402
  [math.AG]}.

\bibitem{arXiv:1103.2673}
\bysame, {\em A framework for tropical mirror symmetry}, {\tt arXiv:1103.2673
  [math.AG]}.

\bibitem{towards-duality}
D.~R. Morrison and M.~R. Plesser, {\em Towards mirror symmetry as duality for
  two-dimensional abelian gauge theories}, Trieste Conference on S-Duality and
  Mirror Symmetry, Nuclear Phys. B Proc. Suppl., vol.~46, 1996, pp.~177--186,
  {\tt arXiv:hep-th/9508107}.

\bibitem{Hori:2000kt}
K.~Hori and C.~Vafa, {\em Mirror symmetry}, {\tt arXiv:hep-th/0002222}.

\bibitem{Hori:2003ic}
K.~Hori, S.~Katz, A.~Klemm, R.~Pandharipande, R.~Thomas, C.~Vafa, R.~Vakil, and
  E.~Zaslow, {\em Mirror symmetry}, Clay Mathematics Monographs, vol.~1,
  American Mathematical Society, Providence, RI, 2003.

\bibitem{hosonokonishi0704}
S.~Hosono and Y.~Konishi, {\em Higher genus {G}romov-{W}itten invariants of the
  {G}rassmannian, and the {P}faffian {C}alabi--{Y}au 3-folds}, Adv. Theor.
  Math. Phys. {\bf 13} (2009) 463--495, {\tt arXiv:0704.2928 [math.AG]}.

\bibitem{arXiv:1011.2350}
M.~Shimizu and H.~Suzuki, {\em Open mirror symmetry for {P}faffian
  {C}alabi--{Y}au 3-folds}, JHEP {\bf 1103} (2011) 083, {\tt arXiv:1011.2350
  [hep-th]}.

\end{thebibliography}
\end{document}